\documentclass[12pt]{iopart}
\usepackage{xcolor}
\usepackage{pifont}
\usepackage{comment}
\usepackage{acronym}
\usepackage{standalone}
\usepackage{academicons}
\usepackage{scalerel}
\usepackage{layouts}
\usepackage{import}
\usepackage{subcaption}
\usepackage{tikz}
\usetikzlibrary{decorations.pathreplacing,angles,quotes,shapes.geometric, arrows,calc}
\usetikzlibrary{svg.path}
\expandafter\let\csname equation*\endcsname\relax 
\expandafter\let\csname endequation*\endcsname\relax 
\usepackage{amsmath}
\usepackage{amssymb}
\usepackage{hyperref}
\usepackage{bibentry}
\usepackage[capitalize]{cleveref}

\usepackage{graphicx}
\usetikzlibrary{decorations.pathreplacing,angles,quotes,shapes.geometric, arrows,calc}
\usepackage{wrapfig, blindtext}
\usepackage{pgfplots}
\pgfplotsset{compat=1.16}

\usepackage[mathlines]{lineno}

\def\prd{\ref@jnl{Phys.~Rev.~D}}
\def\prl{\ref@jnl{Phys.~Rev.~Letters}}

\newcommand{\ndimensional}[1]{$#1$\nobreakdash\discretionary{-}{-}{-}dimensional}

\definecolor{orcidlogocol}{HTML}{A6CE39}
\tikzset{
  orcidlogo/.pic={
    \fill[orcidlogocol] svg{M256,128c0,70.7-57.3,128-128,128C57.3,256,0,198.7,0,128C0,57.3,57.3,0,128,0C198.7,0,256,57.3,256,128z};
    \fill[white] svg{M86.3,186.2H70.9V79.1h15.4v48.4V186.2z}
                 svg{M108.9,79.1h41.6c39.6,0,57,28.3,57,53.6c0,27.5-21.5,53.6-56.8,53.6h-41.8V79.1z M124.3,172.4h24.5c34.9,0,42.9-26.5,42.9-39.7c0-21.5-13.7-39.7-43.7-39.7h-23.7V172.4z}
                 svg{M88.7,56.8c0,5.5-4.5,10.1-10.1,10.1c-5.6,0-10.1-4.6-10.1-10.1c0-5.6,4.5-10.1,10.1-10.1C84.2,46.7,88.7,51.3,88.7,56.8z};
  }
}

\newcommand\orcidicon[1]{\href{https://orcid.org/#1}{\mbox{\scalerel*{
\begin{tikzpicture}[yscale=-1,transform shape]
\pic{orcidlogo};
\end{tikzpicture}
}{|}}}}

\usepackage{hyperref} 

\begin{document}
\title{Generalised gravitational burst generation with Generative Adversarial Networks}

\author{
    J. McGinn \orcidicon{0000-0000-0000-0000},
    C. Messenger \orcidicon{0000-0001-7488-5022},
    I.S. Heng \orcidicon{0000-0000-0000-0000},
    M. J. Williams \orcidicon{0000-0003-2198-2974}
}

\address{SUPA, School of Physics and Astronomy, University of Glasgow, Glasgow G12 8QQ, United Kingdom}
\vspace{10pt}
\begin{abstract}
We introduce the use of conditional generative adversarial networks for generalised gravitational wave burst generation in the time domain. Generative adversarial networks are generative machine learning models that produce new data based on the features of the training data set. We condition the network on five classes of time-series signals that are often used to characterise gravitational wave burst searches: sine-Gaussian, ringdown, white noise burst, Gaussian pulse and binary black hole merger. We show that the model can replicate the features of these standard signal classes and, in addition, produce generalised burst signals through interpolation and class mixing. We also present an example application where a convolutional neural network classifier is trained on burst signals generated by our conditional generative adversarial network. We show that a convolutional neural network classifier trained only on the standard five signal classes has a poorer detection efficiency than a convolutional neural network classifier trained on a population of generalised burst signals drawn from the combined signal class space.
\end{abstract}

%
%
%
%
%

\acrodef{CBC}[CBC]{compact binary coalescence} 
\acrodef{ML}[ML]{machine learning}
\acrodef{AI}[AI]{artificial intelligence}
\acrodef{CNN}[CNN]{convolutional neural network}
\acrodef{GAN}[GAN]{generative adversarial network}
\acrodef{CGAN}[CGAN]{conditional generative adversarial network}
\acrodef{ACGAN}[ACGAN]{auxilliary conditional generative adversarial network}
\acrodef{DCGAN}[DCGAN]{deep convolutional generative adversarial network}
\acrodef{BBH}[BBH]{binary black hole}
\acrodef{SNR}[SNR]{signal-to-noise ratio}
\acrodef{PSD}[PSD]{power spectral density}

\section{Introduction}

%
Gravitational wave (GW) astronomy is now an established field that began with the first detection of a binary black hole merger~\cite{Abbott2016} in September 2015. Following this, the first and second observations runs (O1 and O2) of Advanced LIGO and Advanced Virgo~\cite{Prospects-dets, AdvLIGO, AdvLIGO2, AdvVIRGO} reported several more \ac{CBC} mergers~\cite{Abbott2016a, Abbott2017, Abbott2017a}. On 17th August 2017 a binary neutron star merger was observed alongside its electromagnetic counterpart for the first time, giving rise to multi-messenger GW astronomy \cite{Abbott2017b}. The most recent search for compact binary coalescence, 03a, took place between 1 April 2019 and 1 October 2019 with 39 candidate events reported \cite{GWTC2:2020}.  

%
With these successes and continued upgrades to the detectors \cite{det-upgrades1,det-upgrades2}, further
detections of \acp{CBC} are expected to be commonplace in future advanced
detcetor observation runs. Another group of GW signals that has thus far
been undetected is GW ``bursts". GW bursts are classed as transient
signals of typically short duration ($<$ 1s) whose waveforms are not accurately
modelled or are complex to reproduce. Astrophysical sources for such transients
include: Core collapse supernova~\cite{Fryer_2003}, Pulsar
glitches~\cite{Andersson_2001}, Neutron star post-mergers~\cite{Baiotti_2007}
and other as-yet unexplained astrophysical phenomena. 

%
GW searches for modelled signals use a process called
matched-filtering,~\cite{Owen1998,Usman_2016,sachdev2019gstlal}, where a large template bank of possible
GW waveforms are compared to the detector outputs. For GW bursts that remain unmodelled; there are no
templates available and so matched-filtering is unsuitable for the detection of
these signals.  Instead, detection algorithms like coherent WaveBurst \cite{drago2020coherent} distinguish the signal from
detector noise by looking for excess power contained in the time-frequency
domain and rely on the
astrophysical burst waveform appearing in multiple detectors at similar times.
This is only possible if the detector noise is well characterised and the
candidate signal can be differentiated from systematic or environmental
glitches. 

%
GW burst detection algorithms~\cite{drago2020coherent,Klimenko_2008, Aso_2008} are tested
and tuned using modelled waveforms that have easy to define parameters and share characteristics of real bursts that aim to simulate a GW passing between
detectors. Such waveforms include sine-Gaussians: a
Gaussian modulated sine wave that is characterised by its central frequency and
decay parameter. Bandlimited white noise bursts: white noise that is contained
within a certain frequency range. Ringdowns: which mimic the damped
oscillations after a \ac{CBC} merger. A Gaussian pulse: a short exponential increase then decrease in amplitude and a binary black hole inspiral.
%
%
With the expectation that there will be many more GW detections in the
future, there is a growing need for fast and efficient GW analysis methods
to match the rising number of detections. While still in its infancy, the application of \ac{ML} to GW analyses has already shown great potential in areas of detection~\cite{Gabbard2017,Gebhard_2019,Krastev_2020}, where these techniques have matched the sensitivity of matched filtering for Advanced LIGO and Advanced Virgo GW searches. Similarly, for unmodelled burst search the flexibility of \ac{ML} algorithms has been shown to be a natural and sensitive approach to detection~\cite{2020arXiv200914611S}. Progress has also been made in identifying and classifying detector noise transients or  ``glitches''~\cite{Bahaadini, George_2018,Razzano_2018, 2020arXiv200801262G} and in Bayesian parameter estimation~\cite{gabbard2019bayesian,Chua:2019,green2020gravitationalwave} where \ac{ML} techniques can recover parameters of a GW signal significantly faster than standard methods. Long duration signals like continuous GW require long observing times and therefore have large amounts of data needing to be processed. Current \ac{ML} approaches~\cite{2020PhRvD.102b2005D, 2019PhRvD.100d4009D, 2020arXiv200708207B} are particularly well suited to dealing with this as once trained the searches can be performed quickly.

%
In this work we aim to explore the use of \ac{ML} to generate and interpret
unmodelled GW burst waveforms. Using the generative machine learning
model, \acp{GAN}, we train on five classes of waveforms in the time domain. Working on the assumption that \acp{GAN} construct smooth
 high dimensional vector spaces between their input and output, we can then
explore the space between the five classes to construct new
hybrid waveforms. As all the computationally expensive
processes occur during training, once trained, the model is able to
generate waveforms in fractions of a second and produce waveforms that are difficult to generate with current
techniques. These new varieties of waveforms can then be used to evaluate
detection algorithms, gain new insight into sources of GW
bursts and  allow us to better train our algorithms on a
broader range of possible signals and therefore enhance our detection ability. 

%
This paper is organised as follows. In \cref{ML overview} we introduce the basic ideas of machine learning and discuss the choice of algorithm we used. In \cref{Method} we describe the training data and the details of the model. We present the results of the GAN in \cref{results} and show how unmodeled signals can be produced by interpolating and sampling within latent and class spaces. In \cref{cnn classifier} we show that a \ac{CNN} classifier can be trained to distinguish between sets of our \ac{GAN} generated waveforms from noise only cases. We conclude with a summary of the presented work in \cref{conclusions}.

\section{Machine learning} \label{ML overview}
%
\subsection{Artificial neural networks}

%

\begin{figure}[h!]
\begin{subfigure}[b]{0.30\textwidth}
   \begin{subfigure}[b]{1\textwidth}
   	\centering
	\resizebox{\textwidth}{!}{
 	\includegraphics{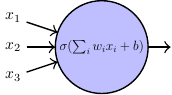}
	}
	\caption{}
	 \label{fig:perceptron}
 \end{subfigure}
 
   \begin{subfigure}[b]{1\textwidth}
  	\centering
	\resizebox{\textwidth}{!}{
\includegraphics{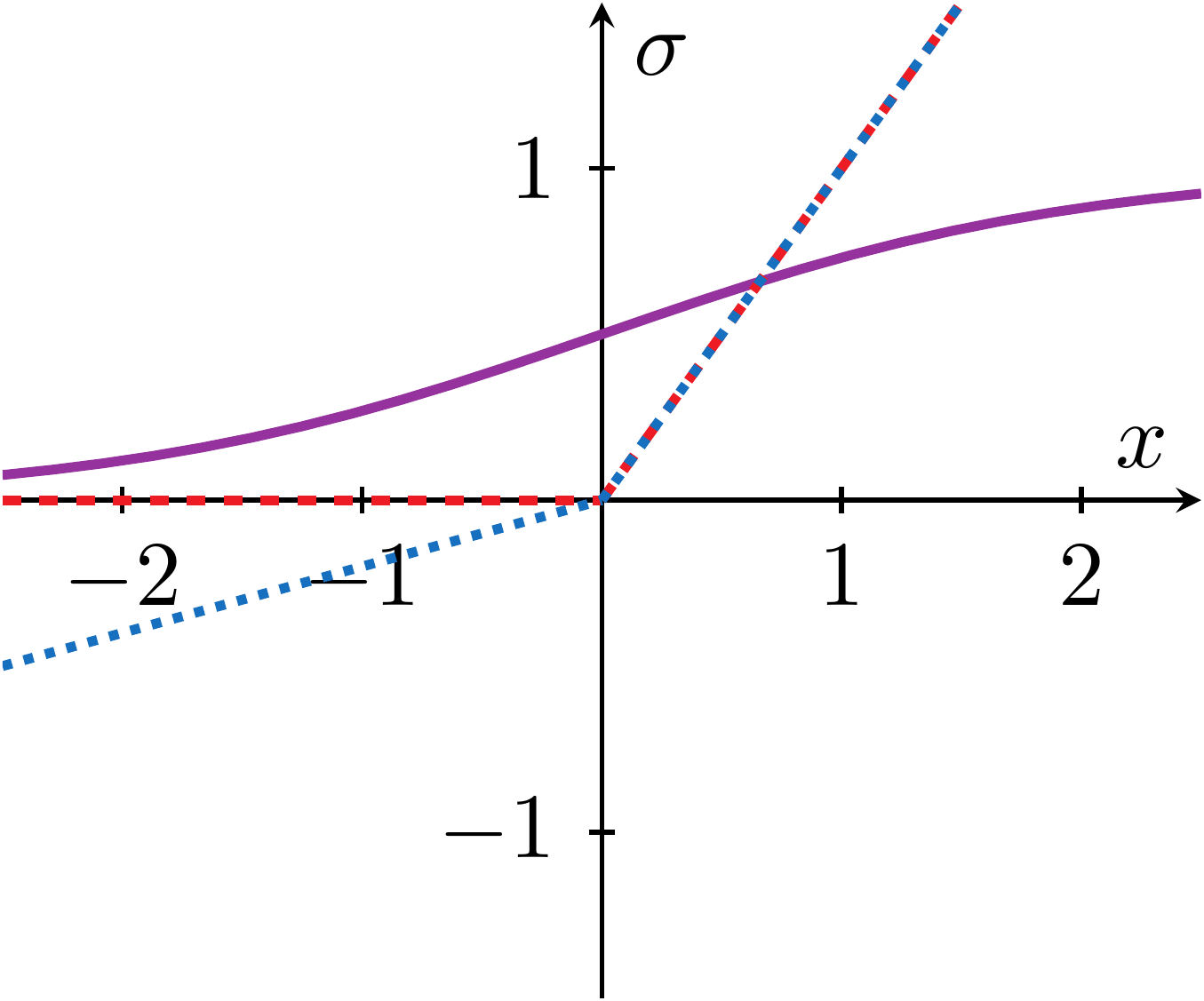}}
\caption{}
\label{fig:activation}
\end{subfigure}
\end{subfigure}
	 \begin{subfigure}[b]{0.70\textwidth}
\centering
\resizebox{\textwidth}{!}{

\includegraphics{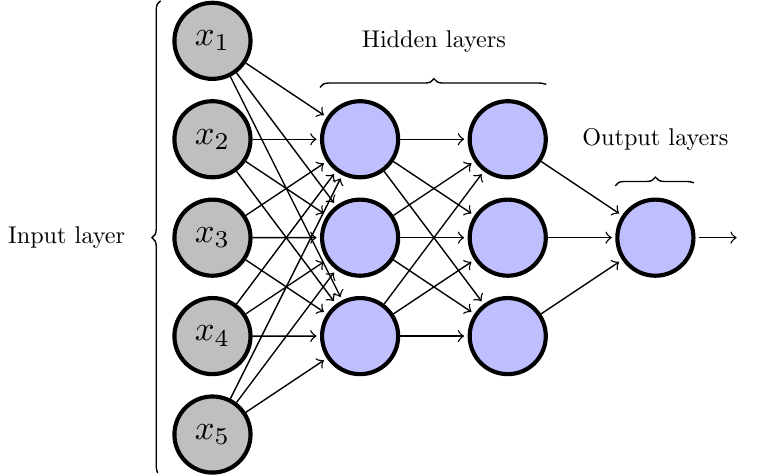}
}
 \caption{}
 \label{fig:network}
 \end{subfigure}

\caption{Neural Networks (a) A single neuron taking a vector of inputs and
returning a single output based on the weights, bias, and activation function
of the network. (b) A selection of activation functions used in this study. The sigmoid (solid purple), rectified linear unit (ReLU)~\cite{relu} (dashed red) and leaky rectified linear unit~\cite{Maas2013RectifierNI} (dotted blue). (c) A an example of a neural network containing two hidden layers that performs a mapping of an input vector to a single output.}
\end{figure}
%
%
\ac{ML} algorithms aim to learn apparent relationships held within given data or `training
data' in order to make accurate predictions without the need for additional
programming. A common approach in \ac{ML} relies on the model learning from
past experience to make decisions on future events. Artificial neural networks are universal function approximators that are built from many single
processing units called neurons. The simplest neural network is the perceptron
layer~\cref{fig:perceptron} which shows a single neuron that takes a vector of real
inputs $x_{i},\ldots, x_{n}$
and maps them to an output according to the linear function, 
\begin{align}
f(x) = \sigma\left(\sum_i w_i x_i + b\right),
\label{eqn:neuron}
\end{align}
where $w$ and $b$ are the weights and bias and $\sigma$ denotes the activation function. The weights are numbers which can be thought of as the strength between connected neurons. The output of a neuron is defined by its activation function which controls how the neuron `fires' depending on its input. Some examples of commonly used activation functions are shown in~\cref{fig:activation}. It is often useful to introduce a bias, $b$, such that the neuron remains inactive above zero but is active when the sum reaches a defined threshold. 

%
A neural network contains many single neurons connected in a layered structure
as shown in~\cref{fig:network}. The activations of the first layer (or
input layer) act as the inputs to the second layer and so on until the output
layer. Multi-layered neural networks have intermediate layers between the input
and output stages dubbed the hidden layers.
%
The output of a single neuron is gives a
prediction that can be compared to the real value through a loss (also known as a
cost) function. The network is trained to minimise this function by updating the weights in the negative
direction of the loss gradient in a process referred to as gradient
descent \cite{ruder2016overview}. The training process for a single layered network is easy to compute as the weights relate directly to the gradient of the loss function the network is trying to minimise. For deeper architectures, the loss is a complicated function of all the weights in all the layers. The backpropagation \cite{Nielsen1992} algorithm acts over the many paths from node to output. It does so in two phases:

\begin{itemize}
\item Forward phase: For one instance of training, the inputs are fed forward through the network using the current weights and the final output is compared to the training labels. The derivative of the loss function is then computed.
\item Backward phase: This phase learns how the gradient of the loss function changes when the weights are varied. Starting at the output node, the algorithm goes backwards through the network (hence the name). The weights that give the steepest descent to the loss function are saved for the next training instance.  
\end{itemize}
This process of updating the weights is repeated until the loss function reaches convergence or a global minimum. As it is impractical to feed the entire data into the network at once, the training is split up into smaller more manageable batches. For this work we train on random samples from the training data and define an epoch as the number of training steps.

\subsection{Convolutional Neural Networks}
%
Convolutional neural networks (CNNs) are designed to work with grid-like input structures that exhibit
strong local spatial dependencies. Although most work with \acp{CNN}
involves image-based data, they can be applied to other spatially adjacent data
types such as time-series \cite{2018arXiv180904356I} and text items \cite{2020arXiv200403705M}. \acp{CNN} are defined by the use of a
convolution operation, a mathematical operation that expresses the amount
overlap between the data. Much like traditional neural networks the convolution operation in this context involves multiplying the input by an array of weights, called a filter or a kernel which is typically smaller in size than the input. The convolution is applied by shifting the kernel over the input, drawing out spatially important features between the
two. The distance by which the grid is shifted is known as
the stride and increasing it reduces the dimension of the output in a process know as downsampling. Alternatively, upsampling the inputs can be achieved using a transposed convolution \cite{dumoulin2016guide}. The output of the convolutional layer is then passed to an activation function and through the next layers. For deep neural networks, techniques like BatchNormalisation \cite{ioffe2015batch} which standardise the inputs to a layer and SpatialDropout \cite{tompson2014efficient} which sever connections between neurons can both help to stabilise learning.

\subsection{Generative Adversarial Networks}
%
%
A subset of deep learning that has seen fruitful development in recent years
are generative adversarial networks \acp{GAN}~\cite{Goodfellow2014}. These unsupervised algorithms learn patterns in a
given training dataset using an adversarial process. The generations from
\acp{GAN} are currently state-of-the-art in fields such as high quality image
fidelity~\cite{brock2018large,karras2019analyzing}, text-to-image
translation~\cite{reed2016generative}, and video
prediction~\cite{liang2017dual} as well as time-series
generations~\cite{esteban2017realvalued}.
%
%
\acp{GAN} train two competing neural networks, consisting of a discriminator
network that is set up to distinguish between real and fake data and a
generator network that produces fake versions of the real data. The generator model performs a mapping from a fixed length vector $\mathbf{z}$ to its
representation of the data. The input vector is drawn randomly from a Gaussian distribution which is referred to as a latent space comprised of latent variables. The latent space is a compressed representation of a data distribution which the generator applies meaning to during training. Sampling points from this space allows the generator to produce a variety of different generations, with different points corresponding to different features in the generations. The discriminator maps its input $\mathbf{x}$ to a probability that the input came from either the training (real) data or
generator (fake).
%
%
During training, the discriminator and generator are updated using batches of data. Random latent vectors are given to the generator to produce a batch of fake samples and an equal batch of real samples is taken from the training data. The discriminator makes predictions on the real and fake samples and the model is updated through minimising the binary cross-entropy function \cite{Goodfellow-et-al-2016}
\begin{equation}
    L = y \log(\hat{y}) + (1 - y) \log(1-\hat{y}),
    \label{eqn:crossentropy}
\end{equation}
where $\hat{y}$ is the network prediction and $y$ is the true output. While training the discriminator, $D$, on real data, we set $y = 1$ and $\hat{y} = D(\mathbf{x})$ which from \cref{eqn:crossentropy} gives $L(D(\mathbf{x}),1) = \log(D(\mathbf{x}))$. While training on fake data produced by the generator, $G$, $y = 0$ and $\hat{y} = D(G(\mathbf{z}))$ and so, $L(D(G(\mathbf{z})),0) = \log(1-(D(G(\mathbf{z}))))$. Since the objective of the discriminator is to correctly classify fake and real data these equations should be maximised, while the goal of the generator should be to minimize these equations. This gives what is know as the \ac{GAN} value function as
\begin{equation}
   \mathop{\text{min}}_{G}  \mathop{\text{max}}_{D} V(D,G) = \mathbb{E}_{\mathbf{x} \sim p_{\text{\text{data}}}(\mathbf{x})} [\text{log} D(\mathbf{x})] + \mathbb{E}_{\mathbf{z} \sim p_{\mathbf{z}}(\mathbf{z})} [\text{log}(1-D(G(\mathbf{z})))],
 \label{equation:GANloss}
 \end{equation}
where $p_{\text{\text{data}}}(\mathbf{x})$ is the distribution of real data and $p_{\text{z}}(\mathbf{z})$ is the latent distribution. 

\subsection{Training stages}
Training a GAN involves updating both the discriminator and generator in stages. First, the discriminator is updated using real instances from the training set. We set the true label $y=1$ and calculate the loss with respect to the predictions $\hat{y}$ via \cref{eqn:crossentropy}. Stochastic gradient descent is used to maximise the loss which has reduced to $L_D(\textrm{real}) = \log(\hat{y})$. The discriminator is then trained on fake instances taken from the generator where we set $y=0$ and maximise $L_D(\textrm{fake}) = \log(1-\hat{y})$. To train the generator, we use a composite model of the generator and discriminator and allow the gradients to flow through this entire model. Following on from what was described before, to train the generator we set y = 0 and minimise $L_{G}$(fake) = $\log(1-\hat{y})$. During early stages of training the generator produces poor generations and so D can easily determine them as fake i.e. $\hat{y})$~0. This leads $L_{G}$ to tend to 0 and we encounter the \textit{vanishing gradient problem}, where the gradients become so small that the weights can no longer be updated. A solution to this problem involves changing the generator loss to maximise $L_{G}$(fake) = $\log(\hat{y})$ or equivalently continue to minimise $L_{G}$(fake) = $\log(1-\hat{y})$ and simply switch the $y$ label to 1. This tweak to the generator loss is called non-saturating generator loss and was reported in the original \ac{GAN} paper \cite{Goodfellow2014}. It was also shown in that paper that if the generator and discriminator can no longer improve, then the discriminator can no longer distinguish between real and fake i.e. $D(x) = \frac{1}{2}$. 

\begin{figure}[h!]
    \centering
    \includegraphics[width=0.85\textwidth]{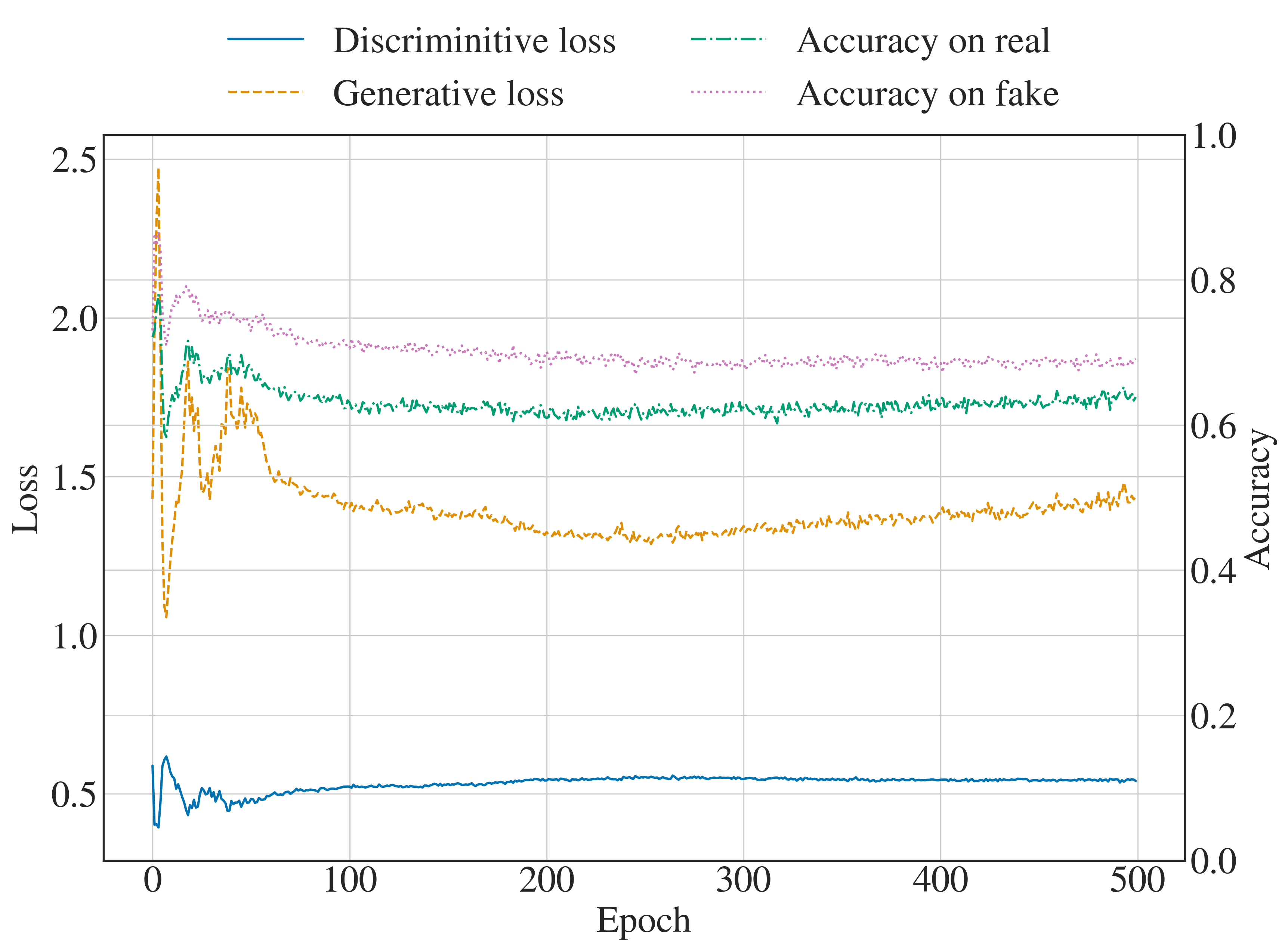}
    \caption{Plot of the discriminator and generator loss and accuracy as a function of epochs. Early in training the losses oscillate as both models attempt to find their equilibrium, after which, both losses vary around a point which signifies stable training. Accuracies on the real and fake data are similar, showing that neither model is stronger than the other.}
    \label{fig:lossplot}
\end{figure}

As \acp{GAN} are trained by updating one model at the expense of the other, they can be hard to train. GANs attempting to replicate complicated structures that do not
have the necessary architecture either struggle to produce results at all or
fall into the common failure mode know as mode collapse; where the generator
produces a small variety of samples or simply memorises the training set. The goal of \ac{GAN} training is to find an equilibrium between the two models, if this cannot be found then it is said that the \ac{GAN} has failed to converge. One way to diagnose problems, such as mode collapse, when training \acp{GAN} is to keep track of the loss and accuracy over time. Loss plots, for example, as seen in \cref{fig:lossplot} can help to identify common failure modes or to check if the \ac{GAN} has indeed converged. Accuracy is another metric that may be used to monitor convergence and is defined as the number of correct predictions made divided by total number of predictions. There is currently no notion of early stopping in \acp{GAN}, instead, training is halted after convergence and by visually inspecting the generations. 

\subsection{Conditional GANs}

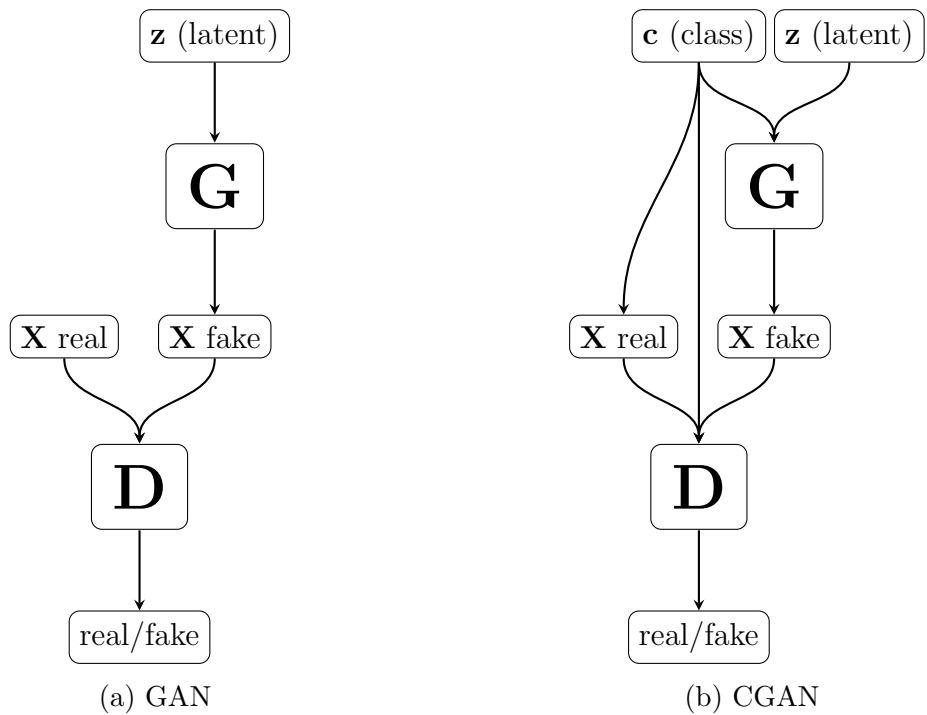
\begin{figure}[h!]
    \begin{subfigure}{.5\textwidth}
     \centering

    
\begin{tikzpicture}[node distance=2cm]

\tikzstyle{zinput} = [rectangle, rounded corners, text centered, draw=black]
\tikzstyle{generator} = [rectangle, rounded corners, text centered, draw=black]
\tikzstyle{X real} = [rectangle, rounded corners, text centered, draw=black]
\tikzstyle{X fake} = [rectangle, rounded corners, text centered, draw=black]
\tikzstyle{X real} = [rectangle, rounded corners, text centered, draw=black]
\tikzstyle{discriminator} = [rectangle, rounded corners, text centered, draw=black]
\tikzstyle{real/fake} = [rectangle, rounded corners, text centered, draw=black]

\tikzstyle{arrow} = [thick,->,>=stealth]

\node (r) [X real] {\textbf{X} real};
\node (f) [X fake, right of = r] {\textbf{X} fake};
\node (G) [generator,above of = f, scale = 2] {\textbf{G}};
\node (z) [zinput] [zinput, above of = G] {\textbf{z} (latent)};
\node (D) [discriminator, below of = f, xshift = -1cm, scale = 2] {\textbf{D}};
\node (rf) [real/fake, below of = D] {real/fake};

\draw [arrow] (z) -- (G);
\draw [arrow] (G) -- (f);
\draw [arrow] (r) edge[out=270,in=90] (D);
\draw [arrow] (f) edge[out=270,in=90] (D);
\draw [arrow] (D) -- (rf);

\end{tikzpicture}
        \caption{GAN}
    \end{subfigure}
    \begin{subfigure}{.5\textwidth}
     \centering
        \begin{tikzpicture}[node distance=2cm]

\tikzstyle{zinput} = [rectangle, rounded corners, text centered, draw=black]
\tikzstyle{generator} = [rectangle, rounded corners, text centered, draw=black]
\tikzstyle{X real} = [rectangle, rounded corners, text centered, draw=black]
\tikzstyle{X fake} = [rectangle, rounded corners, text centered, draw=black]
\tikzstyle{X real} = [rectangle, rounded corners, text centered, draw=black]
\tikzstyle{discriminator} = [rectangle, rounded corners, text centered, draw=black]
\tikzstyle{real/fake} = [rectangle, rounded corners, text centered, draw=black]
\tikzstyle{coutput} = [rectangle, rounded corners, text centered, draw=black]

\tikzstyle{arrow} = [thick,->,>=stealth]

\node (r) [X real] {\textbf{X} real};
\node (f) [X fake, right of = r, xshift = 0cm] {\textbf{X} fake};
\node (G) [generator,above of = f, scale = 2] {\textbf{G}};
\node (z) [zinput] [zinput, above of = G, xshift = 1cm] {\textbf{z} (latent)};
\node (c) [coutput, left of = z] {\textbf{c} (class)};
\node (D) [discriminator, below of = f, xshift = -1cm, scale = 2] {\textbf{D}};
\node (rf) [real/fake, below of = D] {real/fake};

\draw [arrow] (z) edge[out=270,in=90] (G);
\draw [arrow] (c) edge[out=270,in=90] (D);
\draw [arrow] (c) edge[out=270,in=90] (G);
\draw [arrow] (c) edge[out=270,in=90] (r);
\draw [arrow] (G) -- (f);
\draw [arrow] (r) edge[out=270,in=90] (D);
\draw [arrow] (f) edge[out=270,in=90] (D);
\draw [arrow] (D) edge[out=270,in=90] (rf);

\end{tikzpicture}
        \caption{CGAN}
    \end{subfigure}
    \caption{Comparison of the original GAN method and the
conditional-GAN method. $\textbf{G}$ and $\textbf{D}$ denote the generator and discriminator neural networks respectively while $\textbf{X}~\text{real}$ and $\textbf{X}~\text{fake}$ represent samples drawn from the training set and the generated set. For CGANs the training data requires a label denoting
its class that is also fed to the generator which then learns to generate
waveforms based on the input label.}
\label{fig:gan_comparison}
\end{figure}

%
To gain more control over what a \ac{GAN} is able to generate, a conditional variant
of \acp{GAN} named \acp{CGAN}~\cite{cgan} was introduced by feeding in extra
information into the generator and discriminator such as a class label or
attribute label, $\mathbf{c}$. This simple addition has been shown to work well in practice, for instance in image-to-image translation~\cite{isola2016imagetoimage}. We use one-hot encoding to define the classes, that is, each class resides at the corner points of a \ndimensional{5} hyper-cube. For example $\mathbf{c}=[0,1,0,0,0]$ represents the ringdown signal class. The training data and labels are drawn from a joint distribution $p_{\text{data}}(\mathbf{x},\mathbf{c})$, whereas when generating fake data we sample from $\mathbf{c}$ and $p_{\mathbf{z}}(\mathbf{z})$ independently. \cref{equation:GANloss} is modified to include the class labels 
~
\begin{equation}
   \mathop{\text{min}}_{G}  \mathop{\text{max}}_{D} V(D,G) = \mathbb{E}_{\mathbf{x} \sim p_{\text{\text{data}}}(\mathbf{x})} [\text{log} D(\mathbf{x|c})] + \mathbb{E}_{\mathbf{z} \sim p_{\text{z}}(\mathbf{z})} [\text{log}(1-D(G(\mathbf{z|c})))].
 \label{equation:CGANloss}
 \end{equation}
Fig. \ref{fig:gan_comparison} shows the differences in inputs and outputs of a GAN compared with a \ac{CGAN}. We will be using a conditional GAN for this study.

\section{Training data and architecture} \label{Method}
%
\begin{figure}
    \centering
    \includegraphics[width=\textwidth]{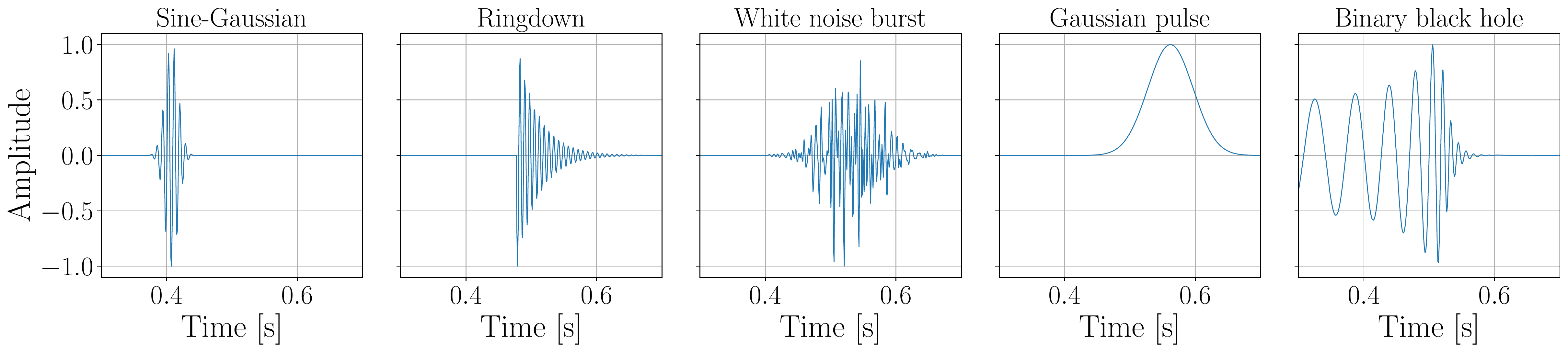}
    \caption{Examples of the five different waveforms that were used in training the \ac{GAN} for this study. Values of the parameters were selected randomly from uniform distributions from \cref{Tab:training_parms}.}
    \label{fig:training_waveforms}
\end{figure}
%
We propose a signal generation scheme using a CGAN trained on burst-like waveforms which we call \texttt{McGANn}\footnote{https://github.com/jmcginn/McGANn}. \texttt{McGANn} is trained on five signal classes which are used to characterise the sensitivity of gravitational wave burst searches (see for example \cite{PhysRevD.100.024017}).

%
\begin{itemize}
\item {\bf Sine-Gaussian}: $h_{\text{SG}}(t) = A \exp\left[ - (t-t_{0})^2 /
\tau^2 \right] \sin (2 \pi f_0 (t-t_0) + \phi)$, a sine wave with a Gaussian envelope characterised by a central frequency $f_0$, amplitude $A$, time of arrival $t_{0}$ and phase $\phi$ which is uniformly sampled between $[0, 2\pi]$.
\item {\bf Ringdown}: $h_{\text{RD}}(t) = A \exp \left[-{(t-t_0)} / {\tau}
\right] \sin(2 \pi f_0 (t-t_0) + \phi)$, with frequency $f_0$ and duration $\tau$, amplitude $A$, time of arrival $t_{0}$ and phase $\phi$ which is uniformly sampled between $[0, 2\pi]$. 
\item {\bf White noise bursts}: $h_{\text{WN}}(t_j) = Ag_j\exp\left[ -
(t-t_{0})^2 / \tau^2 \right]$ where $g_j$ are drawn from a zero mean unit
variance Gaussian distribution with a Gaussian envelope of duration $\tau$.
\item {\bf Gaussian pulse}: $h_{\text{GP}}(t) = \exp(-t^2 / \tau^2)$ with
duration parameter $\tau$.
\item {\bf Binary black hole}: Simulated using the IMRPhenomD
waveform~\cite{Khan_2016} routine from LALSuite~\cite{lalsuite} which models
the inspiral, merger and ringdown of a \ac{BBH} waveform. The component masses
lie in the range of [30,70] $\textrm{M}_{\odot}$ with zero spins and we fix
$m_1>m_2$. The mass distribution is approximated by a power law with
index of 1.6~\cite{Abbott_2019}. The inclinations are drawn
such that the cosine of the angles lies uniformly in the range [-1,1] and we only use the plus polarisation.
\end{itemize}
The location of the peak amplitude of the waveforms (corresponding to the
mid-points of all but the ringdown and \ac{BBH} classes) are randomly drawn from a uniform distribution to
be within $[0.4, 0.6]$ seconds from the start of the 1 second time interval and all
training waveforms are sampled at 1024 Hz.  The parameter prior ranges are
defined in~\cref{Tab:training_parms} and a sample of training waveforms are shown in \cref{fig:training_waveforms}. All training data is rescaled such that their amplitudes peak at 1.
\begin{table}[!h]
\centering
\caption{The parameters used in generating the training data. Each parameter is drawn uniformly in the below ranges.}
\begin{tabular}{@{} l l l l l l }
\br
\hline
 Waveform & Central frequency  & Decay & Central time epoch & Mass range \\
 & (Hz) & (s) & (s) & ($\textrm{M}_{\odot}$) \\
\mr
Sine-Gaussian & 70 - 250 & 0.004 - 0.03 & 0.4 - 0.6 & -  \\  
Ringdown & 70 - 250 & 0.004 - 0.03 & 0.4 - 0.6 & - \\
white noise burst & 70 - 250 & 0.004 - 0.03 & 0.4 - 0.6 & -  \\
Gaussian pulse & - & 0.004 - 0.03 & 0.4 - 0.6 & -  \\
BBH & - & - & - & 30 - 70  \\
 \br
\end{tabular}\\
\label{Tab:training_parms}
\end{table}
\normalsize

With the exception of the binary black hole waveforms, the signal classes described above are analytic proxy waveforms to gravitational wave signals expected from various burst gravitational wave sources. For example, numerical relativity simulations show that rapidly rotating stellar core collapse emit gravitational waves that look like sine gaussians with small time constants (low Q). Additionally, gravitational waves from hyperbolic black hole encounters look very similar to sine gaussians and gaussian pulses. Ringdown signals can be emitted by excited isolated neutron stars, for example, after a pulsar glitch and white noise burst signals mimic the stochastic nature of gravitational wave signals emitted by neutrino driven stellar core collapse.

\subsection{Architecture details}
%
%
%
Neural networks and subsequently \acp{GAN} have multiple parameters a developer
can tune when designing the model and these are referred to as hyperparameters.
The final network design used in this work was developed through trial and
error and the initial designs were influenced by the available literature. We found
that the \ac{GAN} performed better with both networks having the same number of
layers and neurons which encourages even
competition between the generator and discriminator.  After tuning the multiple
hyperparameters (see \cref{Tab:gan_training_parms}), the \ac{GAN} was trained on
$10^5$ signals  drawn from a categorical
distribution with equal propabilities for each class of
sine-Gaussian,
ringdown, white noise bursts, Gaussian pulse and \acp{BBH}.

The design of the networks is influenced by \cite{Radford2015} in which they use a \ac{DCGAN} architecture. The generator model is fully convolutional, upsampled using strided transposed convolutions with BatchNormalisation in the first layer and ReLU activations throughout with the exception of a linear activation for the output layer. The use of a linear activation guarantees the output can have negative and positive outputs. Each transposed convolutional layer uses a kernel size of 18 and stride of 2. The discriminator network mirrors that of the generator without batch normalization, using LeakyReLU activations, SpatialDropout, and a 2-stride convolution for downsampling. The discriminator output is a single node with sigmoid activation that can be interpreted as a probability of the the signal being real and both models are trained with binary cross entropy \cref{eqn:crossentropy}. The full architecture description can be seen in~\cref{Tab:gan_training_parms}.

All models were designed with the Python Keras library \cite{chollet2015keras} and TensorFlow \cite{tensorflow2015-whitepaper} and trained on a GeForce RTX 2080 Ti GPU. We train the networks for 500 epochs which takes $\mathcal{O}(10)$ hours and save the model at each epoch. We choose an appropriate model by visually inspecting the generations at a point of convergence on the loss plot. 
\section{Results} \label{results}
%
Given a \ndimensional{100} vector drawn from a normally distributed latent space and a one-hot encoded class
label, the GAN is able to generate burst-like waveforms generalised from the
training set. We set out by describing the quality of generated waveforms and
how they compare to the training set. We then explore the structure of the
latent and class spaces by interpolating between points in these spaces. We
test three methods of sampling from the class space that can be used to generate new signals composed of weighted elements of each training class.

\begin{figure}[!h]
    \centering
    \includegraphics[width=\textwidth]{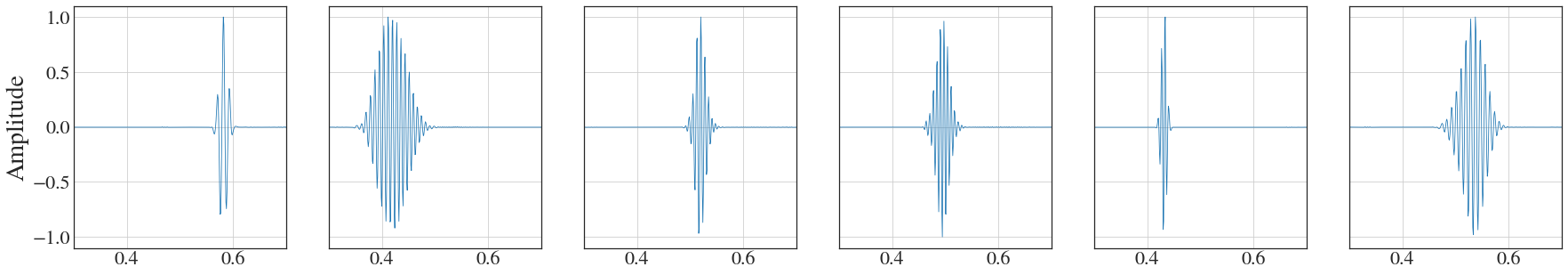}
    \includegraphics[width=\textwidth]{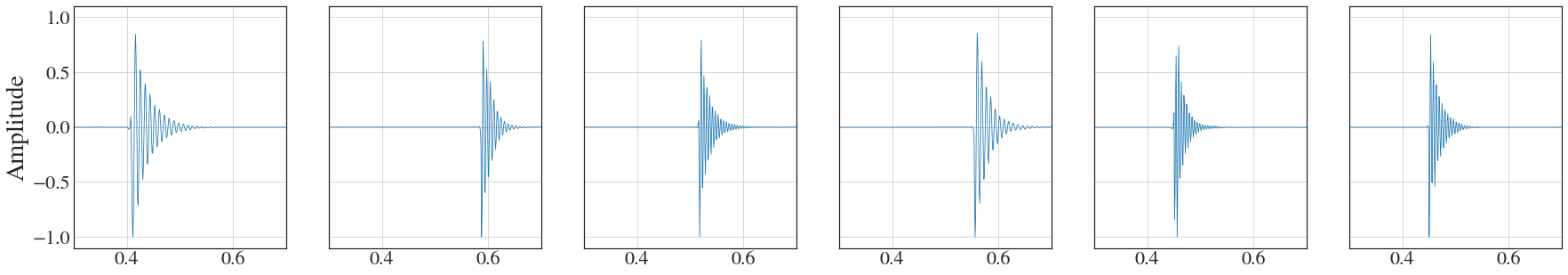}
    \includegraphics[width=\textwidth]{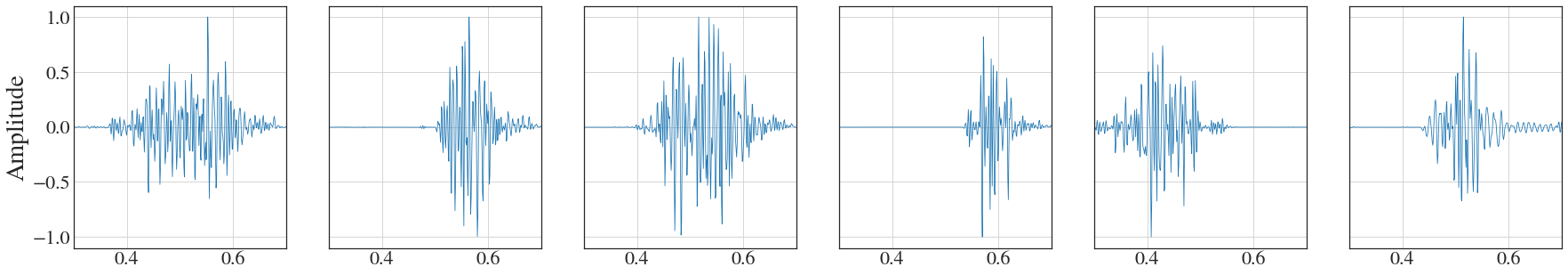}
    \includegraphics[width=\textwidth]{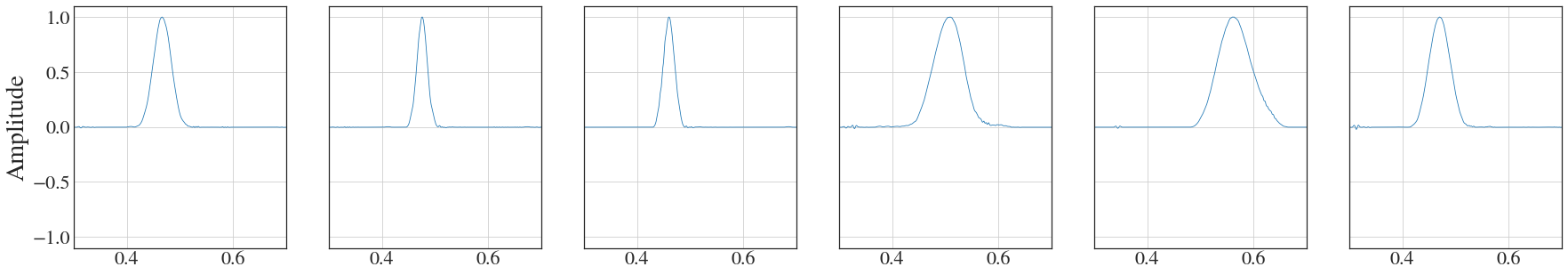}
    \includegraphics[width=\textwidth]{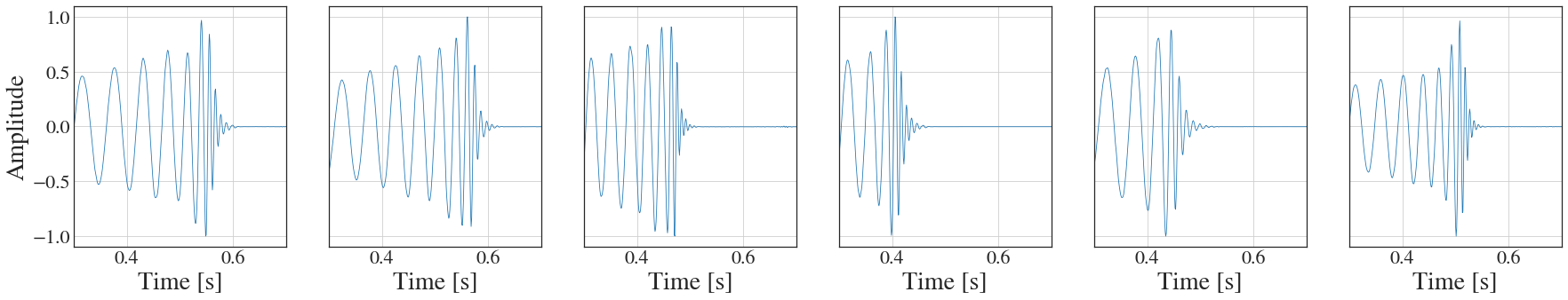}
    \caption{\ac{GAN} Generated waveforms plotted as a function of time. The latent space inputs for each panel are randomised and each row is assigned one of the five class vectors. By row: sine-Gaussian, ringdown,
white noise burst, Gaussian pulse, binary black hole merger. For ease of viewing, the $x$-axis for all panels spans the mid 50\% of the output range.}
\label{fig:gen_signals} 
\end{figure}
\subsection{Known class signal generation}
In \cref{fig:gen_signals} we show conditional signal generations using our generator network. We can see the generations capture the main aspects of each signal class and appear as though they could have plausibly come from the training set. We can also see that the model has learned the overall characteristics of the five training classes and is able to disentangle each class and associate them with the conditional input. Additionally, as the latent variable changes we see indirect evidence of variation within the parameter space for a given class. For instance \cref{fig:gen_signals} shows how signals vary in frequency, central epoch, decay timescale, and phase. The \acp{GAN} ability to generate a variety of signals for various latent space input indicates stable training and no mode collapse. 
\begin{figure}[!h]
    \centering
    \includegraphics[width=\textwidth]{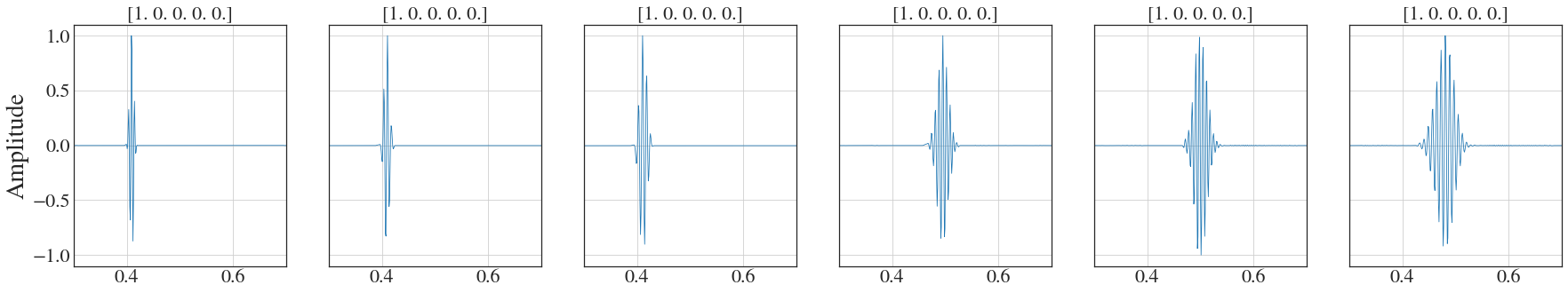}
    \includegraphics[width=\textwidth]{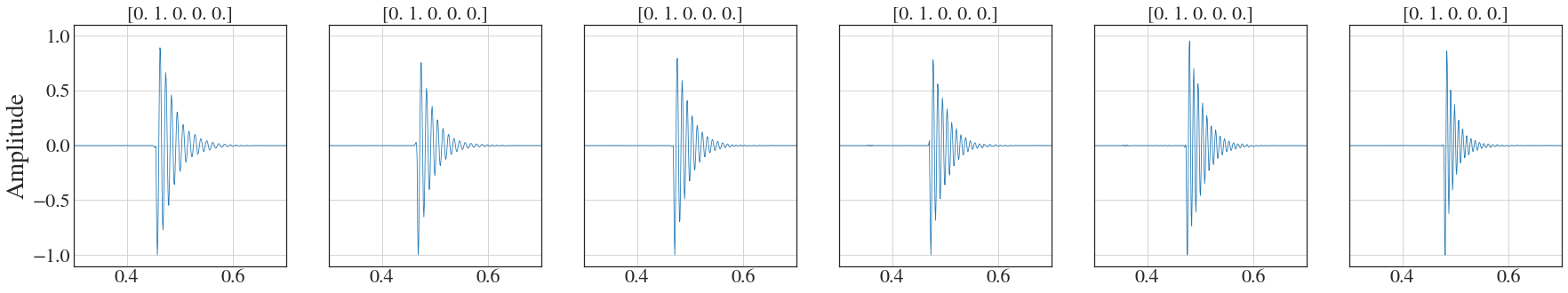}
    \includegraphics[width=\textwidth]{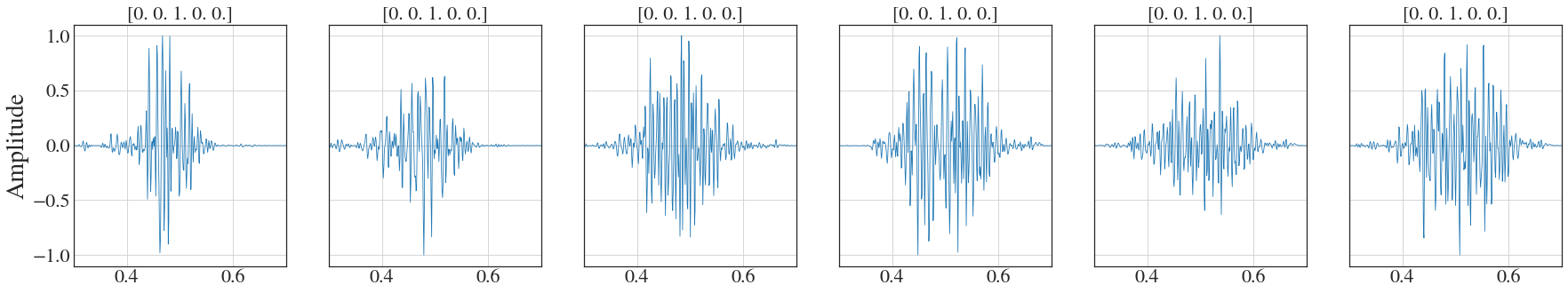}
    \includegraphics[width=\textwidth]{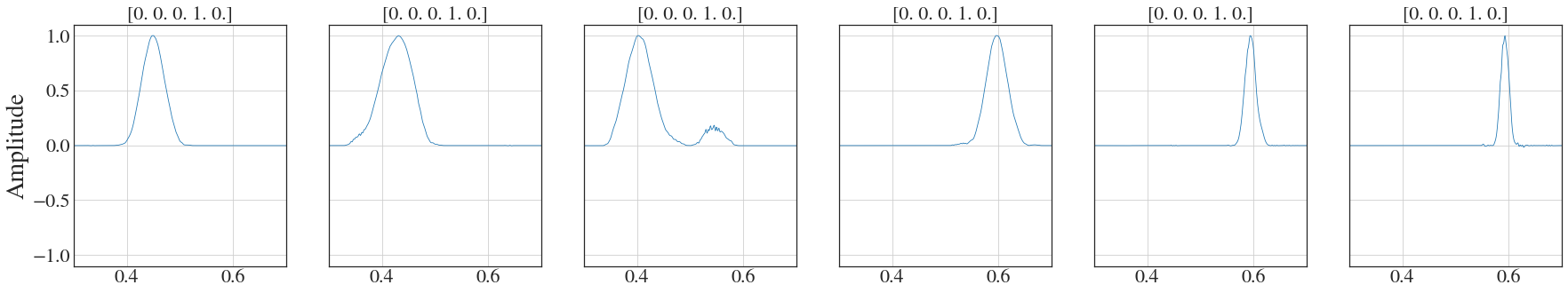}
    \includegraphics[width=\textwidth]{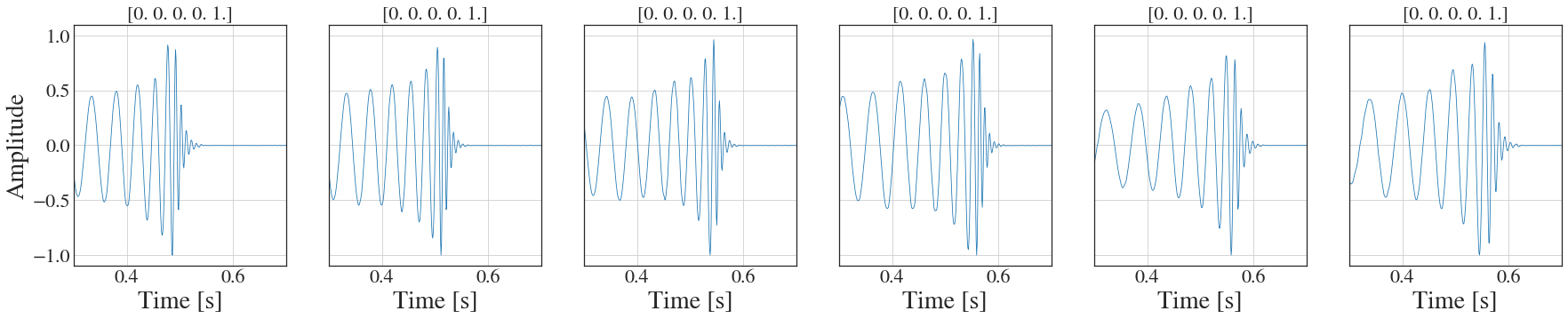}
    \caption{\ac{GAN} generated interpolated waveforms plotted as a function of time showing latent space interpolations. For each interpolation two different points were randomly chosen in the latent space and represent the first and last panels in each row. The panels between represent signals generated using linearly interpolated vectors between these two points. Each row keeps its class vector constant throughout the latent space interpolation. By row: sine-Gaussian, ringdown, white noise burst, Gaussian pulse, binary black hole merger. For ease of viewing, the $x$-axis for all panels spans the mid 50\% of the output range.}
    \label{fig:z_interp}
\end{figure}
\subsection{Interpolation within the latent space}
%
%
 We have shown that the generator produces quality signals and that the model responds well to randomly sampled Gaussian latent vectors. We now assume that during training the generator has learned a mapping from a Gaussian latent space to the signal space and that this mapping is a smooth function of the underlying latent space. To verify this, we fix the class vector input and linearly interpolate between two randomly chosen points in the latent space (different for each class). In \cref{fig:z_interp} we show the generated waveforms, with the class vectors held constant along each row. We can see that each plot shows plausible waveforms suggesting that the generator has constructed a smooth traversable, space. We note that the relationship between the latent space location and the physical signal parameters is intractable, and hence the initial and final latent space locations (moving left to right in \cref{fig:z_interp}) simply represent random possible signals learned from the training set prior. During training the network should have learned how to smoothly represent the underlying features of a signal as a function of latent space location. For example, the linearly interpolated transition through the latent space for the Gaussian pulse signal shows a shift to earlier epoch and larger decay timescale. In contrast, the transition for the ringdown signal appears to pass through a localised region of latent space consistent with higher central frequency. 
\subsubsection{Interpolation between pairs of classes}

\begin{figure}[!h]
    \centering
    \includegraphics[width=\textwidth]{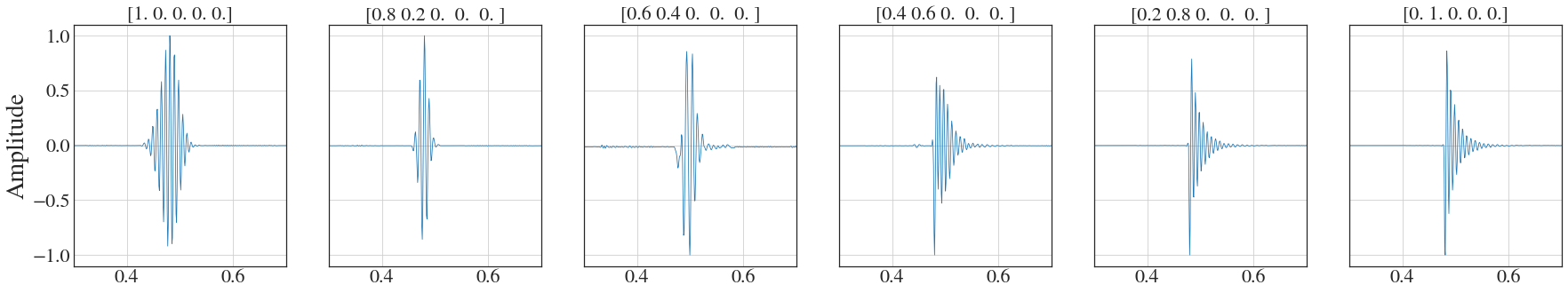}
    \includegraphics[width=\textwidth]{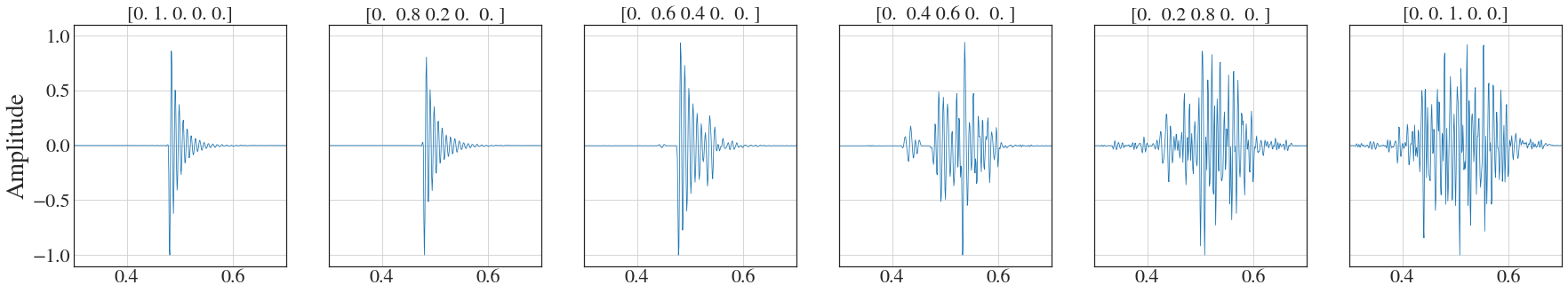}
    \includegraphics[width=\textwidth]{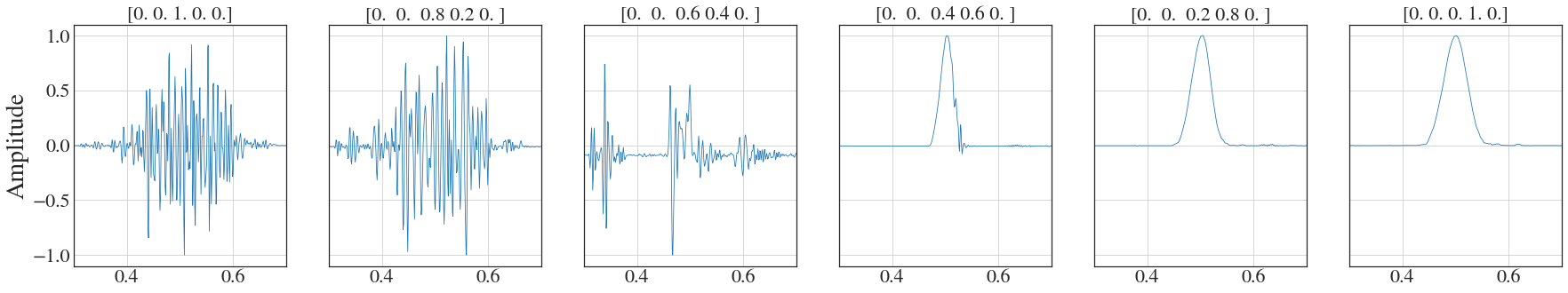}
    \includegraphics[width=\textwidth]{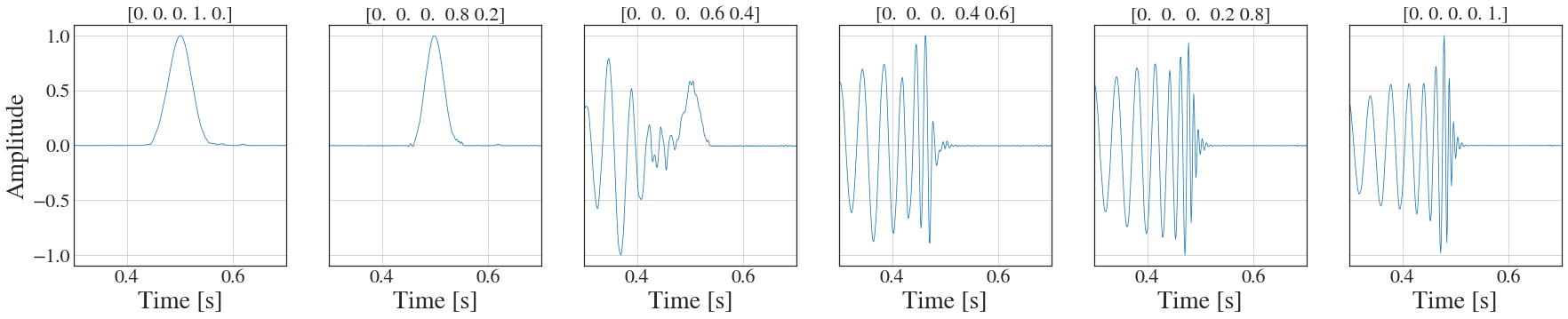}
    \caption{\ac{GAN} generated class interpolated waveforms as a function of time showing class space interpolations. A single latent space vector is used for all generations and is chosen randomly in the latent space. Each row shows generations using linearly interpolated classes as inputs to the generator. By row top to bottom: Sine-Gaussian to ringdown, ringdown to white noise burst, white noise burst to Gaussian pulse, Gaussian pulse to BBH.}
    \label{fig:c_interp}
\end{figure}

While the \ac{GAN} is trained on distinct one-hot encoded classes, we may test arbitrary points in the \ndimensional{5} class space to produce indistinct or hybrid waveforms. In order to explore the class space, in \cref{fig:c_interp} we show results where the latent vector is held constant but we instead linearly
interpolate within the one-hot encoded class space between pairs of the well-defined training class locations. In this scenario we highlight that the \ac{GAN} has not yet probed this intermediate class space during its training and therefore we are reliant on the generator having learned any underlying class space relationships between the 5 training classes. The results show that for each case that the generated signals show distinct characteristics of the respective class pairs at most stages of the transition. We note that transitions in some cases appear to be rather abrupt, e.g., between the Gaussian pulse and the \ac{BBH}, and that this feature, whilst not uncommon, is a strong function of the random latent space location. 
\subsection{General points within the class space}
We have shown that the \acp{GAN} latent space and class space have structure that can be navigated via interpolation between pairs of locations within each respective space. Taking a step further, we can sample from the class space in novel ways to create new inputs for the generator. These new points are categorised by the method used to sample from the class space. The methods we use are divided into the following: 
\begin{itemize}
\item {\bf Vertex}: Points that lie at the corners of the \ndimensional{5} class space. These class space locations are equivalent to the training set locations and are our closest generated representation of the training set.
\item {\bf Simplex}: This class vector we define as uniformly sampled points on a simplex, which is a generalization of a triangle in $k$-dimensions. We sample uniformly on the $k=4$ simplex that is embedded in the \ndimensional{5} class hyper-cube. In practice we use the equivalent of sampling points from a $k=4$ Dirichlet distribution. It is useful to think of the simplex as the hyper-plane that intersects all 5 training classes. It is a subspace of the Uniform method.  
\item {\bf Uniform}: Each of the entries in the class vector is sampled from a uniform distribution $\text{U}[0,1]$. This is equivalent to sampling uniformly within the \ndimensional{5} one-hot encoding hyper-cube.
\end{itemize}

The vertex points are the most straightforward where one element of the class vector contains one and the other elements are zero. These points are equivalent to the class vectors that the \ac{GAN} is trained on e.g., $\mathbf{c} = [1,0,0,0,0]$ would correspond to a sine-Gaussian generation. Uniform class vectors with each element sampled from a uniform distribution are equivalent to a random draw from a \ndimensional{5} hyper-cube. Uniformly sampling generates class space locations up to a maximum distance of unity from the closest class e.g. $[0,0,0,0,0]$ is of distance 1 away from all classes. For simplex class vectors, we sample from the simplest hyper-surface that intersects all the classes and has a symmetry such that no training class location (any vertex) is favoured over any other. For our \ndimensional{5} case this corresponds to a 4-simplex manifold. Sampling from the simplex can be seen as sampling from the simplest space that spans between the training classes.   

\begin{figure}
    \centering
    \includegraphics[width=\textwidth]{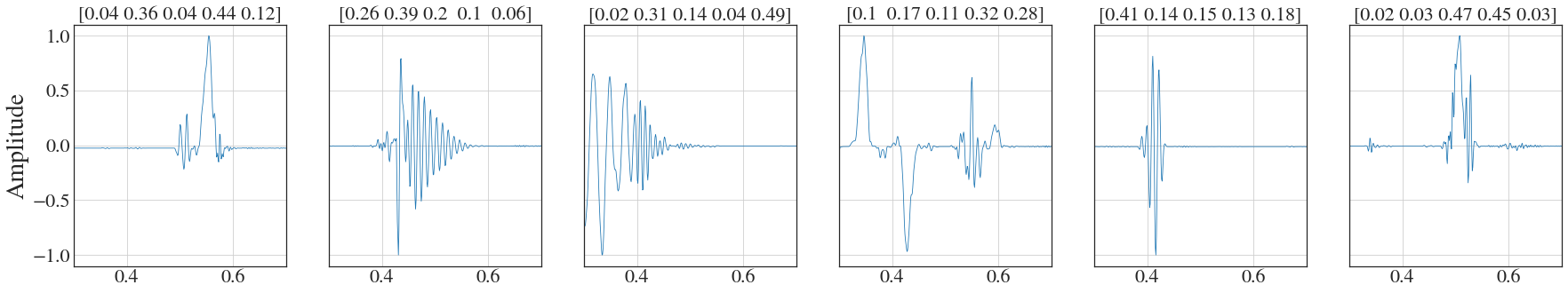}
    \includegraphics[width=\textwidth]{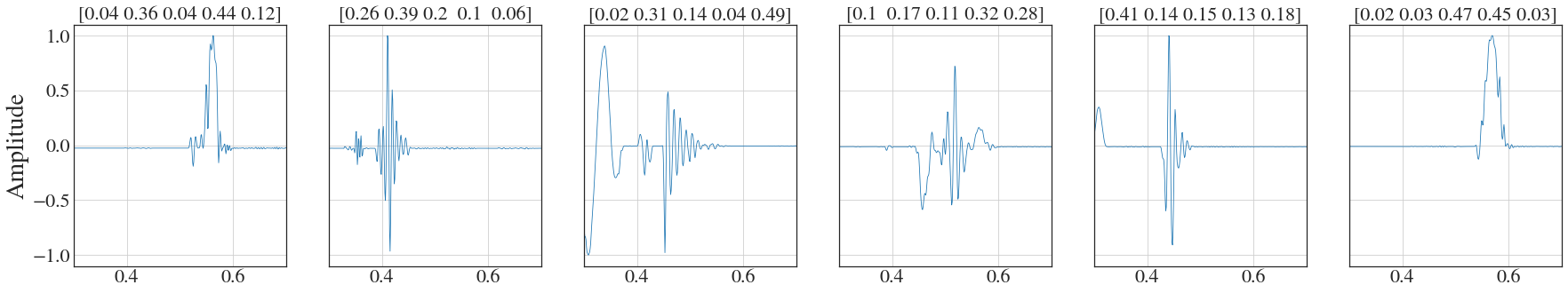}
    \includegraphics[width=\textwidth]{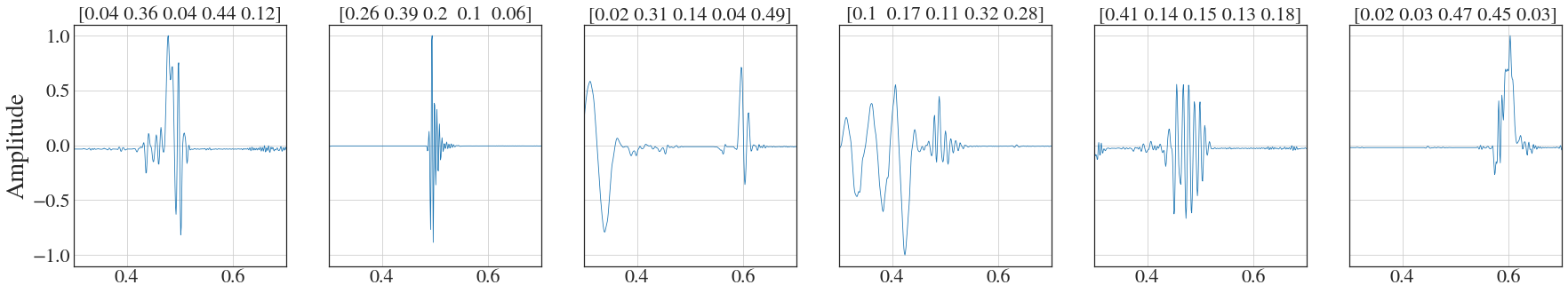}
    \includegraphics[width=\textwidth]{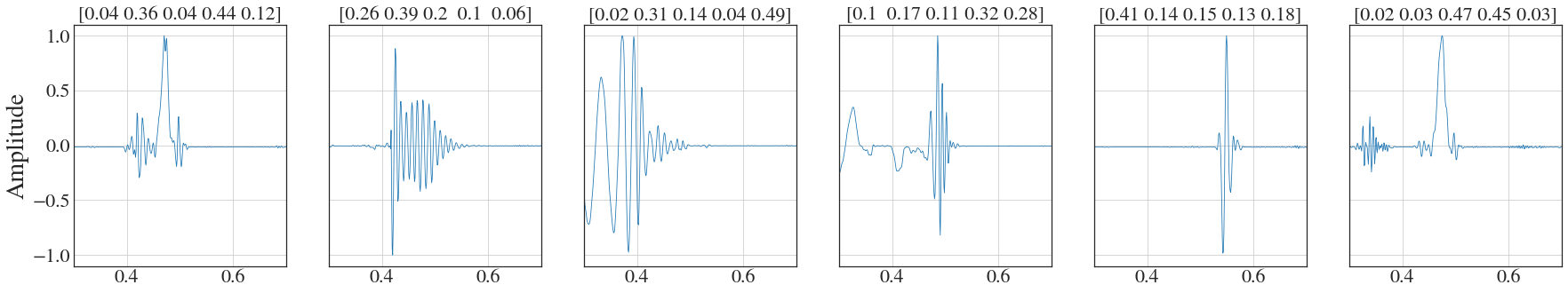}
    \includegraphics[width=\textwidth]{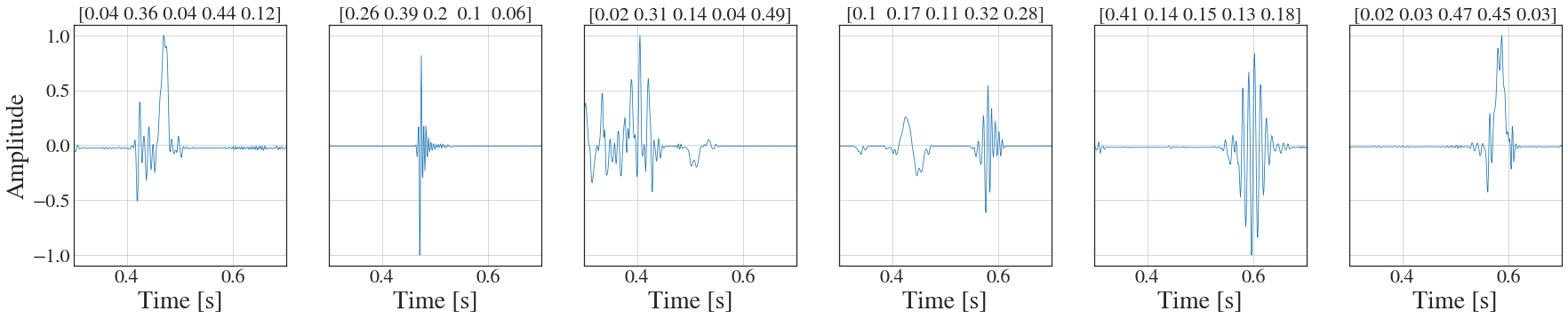}
    \caption{GAN generations where the class vectors are sampled from the \ndimensional{4} plane (simplex) intersecting all training classes. Latent space locations for all signals are drawn randomly from a \ndimensional{100} Gaussian distribution and the signals are then re-scaled such that they have maximum absolute amplitude at unity. The class label for each generation is shown above each panel.}
    \label{fig:simplexd_samples}
\end{figure}

\begin{figure}
    \centering
    \includegraphics[width=\textwidth]{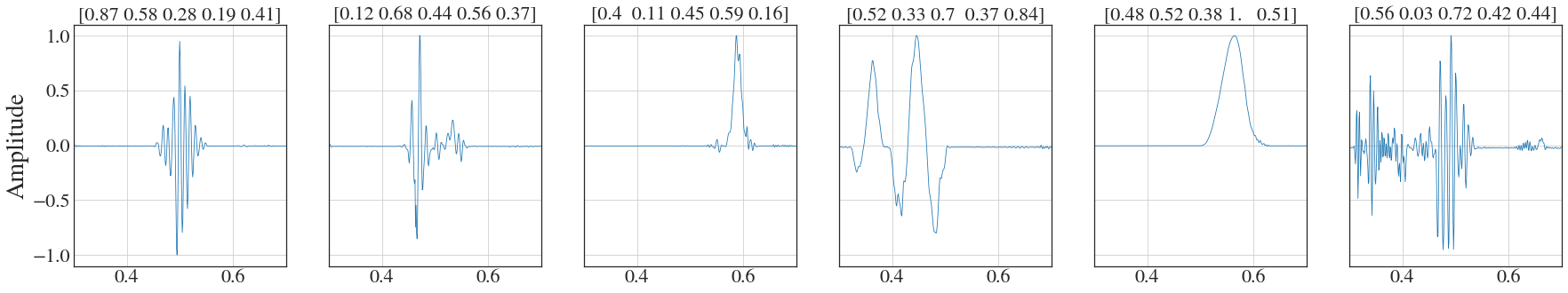}
    \includegraphics[width=\textwidth]{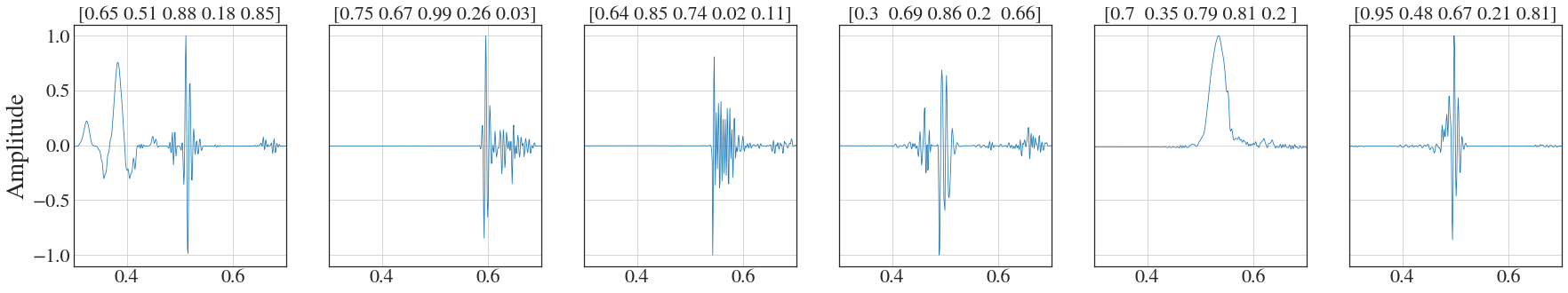}
    \includegraphics[width=\textwidth]{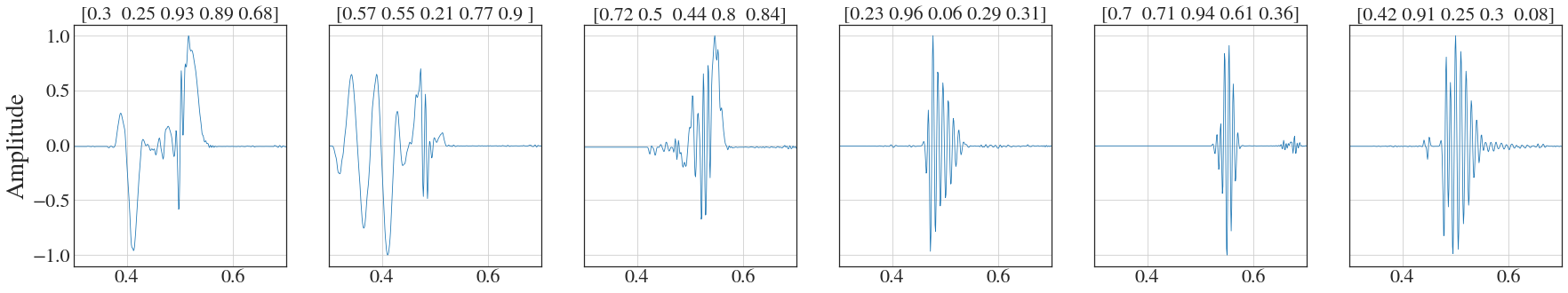}
    \includegraphics[width=\textwidth]{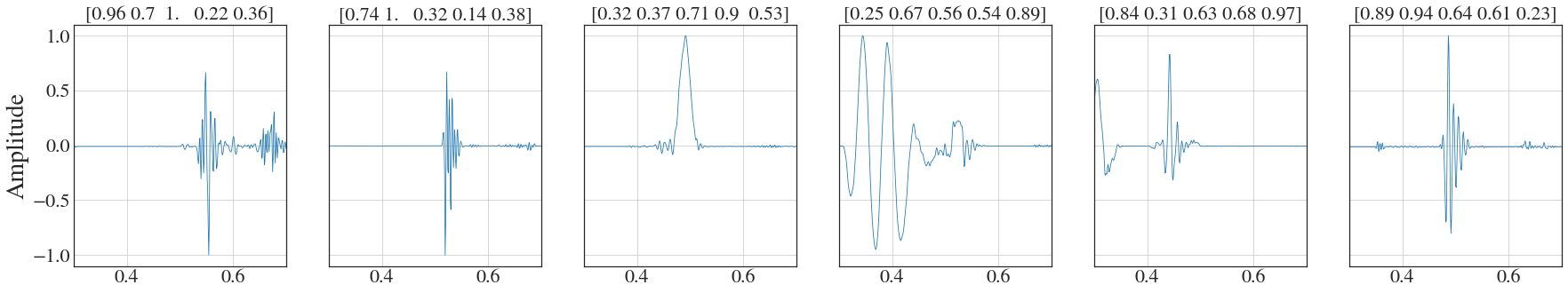}
    \includegraphics[width=\textwidth]{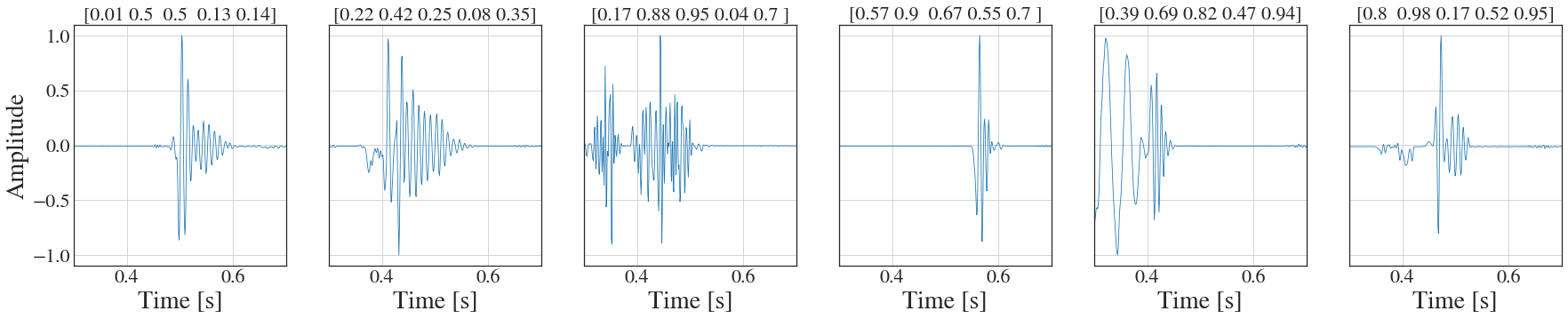}
    \caption{GAN generations where the class vectors are sampled uniformly in the hyper-cube class space. Latent space locations for all signals are drawn randomly from a \ndimensional{100} Gaussian distribution and the signals are then re-scaled such that they have maximum absolute amplitude at unity.}
    \label{fig:uniform_samples}
\end{figure}

In \cref{fig:simplexd_samples} we show generations conditioned on class vectors drawn randomly from the 4-simplex. There are large variations in the signals with some having characteristics strongly resembling the training classes, although this can be partially explained through the random draws from the simplex as there is finite probability that one class entry will dominate over the others (i.e., the class space location is close to a vertex). For instance the generations that look more like sine-Gaussians than hybrid waveforms generally have a larger value placed in the first class space element than others. Similarly \cref{fig:uniform_samples} shows generations conditioned on class vectors drawn uniformly in the unit hyper-cube. These types of generations tend to exhibit more noise and some tend to be generated with very low amplitude prior to being re-scaled to have maximum amplitude of unity. Both methods of generating hybrid waveforms, however, do produce signals that appear to share characteristics from the training set but still be distinct in signal morphology. Upon inspection of a larger collection of waveform generations from both methods we do see a tendency for the uniform hyper-cube approach to generate a wider variety of hybrid waveforms that are more visually distinct from the training set. This is to be expected given that the simplex class space is a subset of the hyper-cube and does not explore regions of the class space as far from the training set vertices.

\section{CNN burst classifier} \label{cnn classifier}
In this section we develop a basic search analysis using a \ac{CNN} in order to compare the sensitivity of such a search using different \ac{GAN} generated waveforms in additive noise. We train a \ac{CNN} to perform simple classification and to distinguish between two classes: signals in additive Gaussian noise and Gaussian noise only. We are primarily interested in the relative sensitivity as a function of the types of waveforms used for training the network. We are also interested in how these differently trained networks perform when applied to data from waveform generations not used in the training process.

\subsection{Noisy datasets} 
\begin{figure}[ht!]
    \centering
    \includegraphics[width=\textwidth]{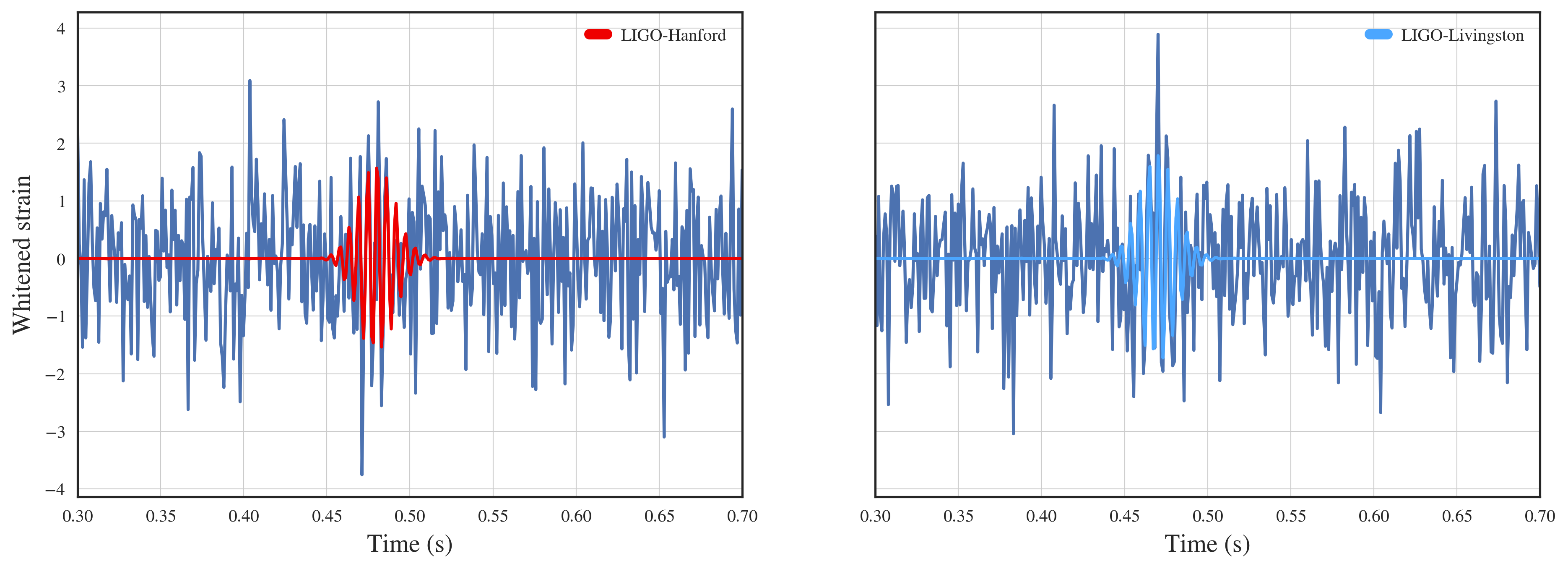}
    \caption{Example of \ac{CNN} training data showing a whitened noisy (dark blue) and noise-free (red, light blue) sine-Gaussian time-series as seen by Hanford (left) and Livingston (right) detectors. This signal has network $\text{SNR}=8$.}
     \label{fig:cnn_training}
\end{figure}

We use three classes of waveforms: vertex, uniform, and simplex cases generated using our \ac{GAN} method. We then construct noisy time-series data from each waveform representing measurements from the 2 LIGO detector sites, Hanford (H1) and Livingston (L1). For each training set we generate $2\times 10^5$ signals and apply antenna responses and sky location dependent relative time delays using routines provided within LALsuite \cite{lalsuite}. The generated waveforms are used to represent the plus-polarisation component of signal only and the polarisation angles are drawn uniformly in the range $[0,2\pi]$ and sky positions are sampled isotropically. Time delays between detectors are computed relative to the Earth's centre. All of the training data used is whitened using the Advanced LIGO design sensitivity \ac{PSD} \cite{designcurve,observing-prospects}, such that there is equal noise power at each frequency. Signal network \acp{SNR} is drawn uniformly in the range $[1,16]$ and is controlled by an amplitude scaling applied to the waveform. Each 1 second duration time-series input to the \ac{CNN} is represented by a \ndimensional{1} 1024 sample vector with 2 channels representing each detector. Example time-series from each detector for a single signal are shown in \cref{fig:cnn_training}. The network is trained to be able to identify whether or not a measurement contains a signal and therefore 50\% of the training data have time-series containing signals and 50\% have only noise. We randomly divide the data into the 3 standard sets (training, validation, and test data) where 40\% is used for training, 10\% used for validation, and 50\% is used for testing in order to achieve suitably low false-alarm probability of $10^{-3}$ . For the Uniform and Simplex datasets samples are drawn uniformly from their respective spaces. For the vertex dataset the 5 different vertex locations in class space are sampled with equal probability.  

\subsection{CNN architecture}
In this approach the inputs to the \ac{CNN} are 1024 sample time-series (with two channels representing each detector output) which are passed through a series of four convolutional layers, onto two fully connected or ``Dense'' layers and finally to a single output neuron which represents the probability that a signal is present within the noise. We used dropout in the final dense layer and used a selection of different activation functions including the swish activation \cite{ramachandran2017searching} which improved overall performance, and a sigmoid activation for the output layer. We used binary cross-entropy \cref{eqn:crossentropy} as the loss function and Adam as an optimizer with learning rate set to $10^{-3}$. In total we train three separate \acp{CNN} on the vertex, uniform and simplex datasets respectively. In each case the networks share the same architecture and hyperparameters which are defined in \cref{Tab:cnn_training_parms}.

\subsection{CNN results}

\begin{figure} 
\centering
\begin{subfigure}{1\textwidth}
\centering
\includegraphics[scale=0.85]{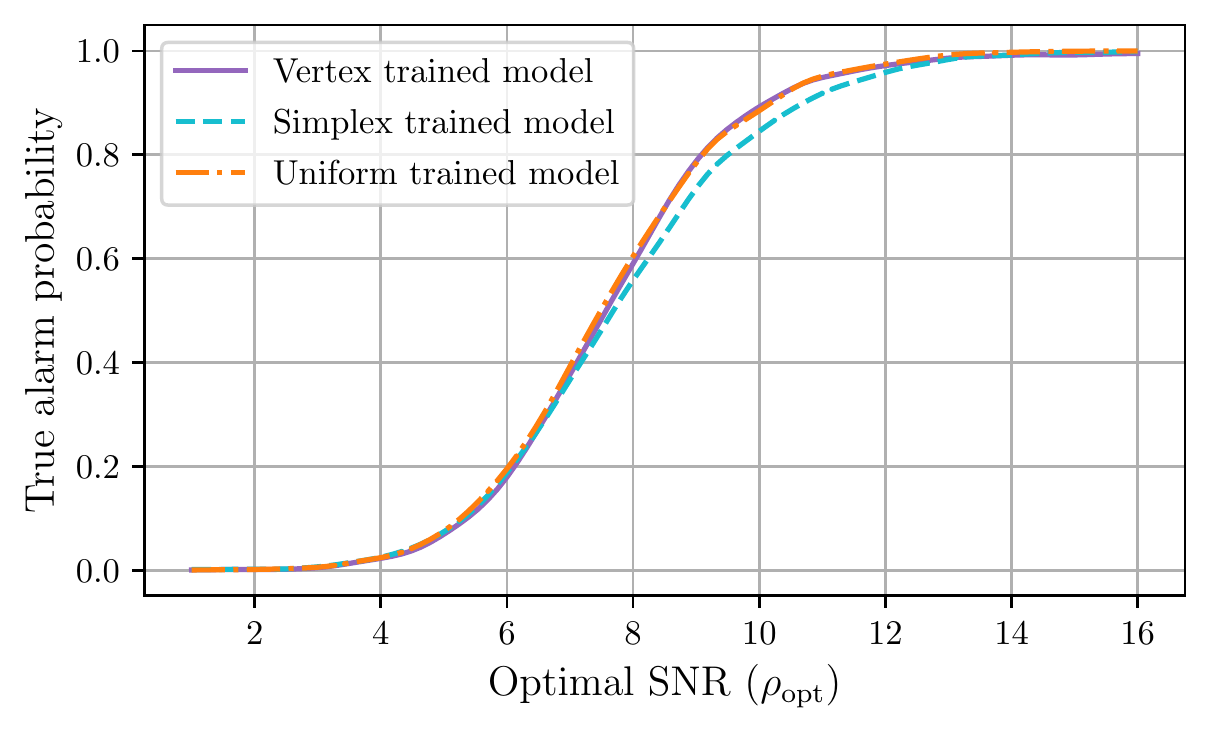}
\caption{Tested on vertex}
\end{subfigure}

\begin{subfigure}{1\textwidth}
\centering
\includegraphics[scale=0.85]{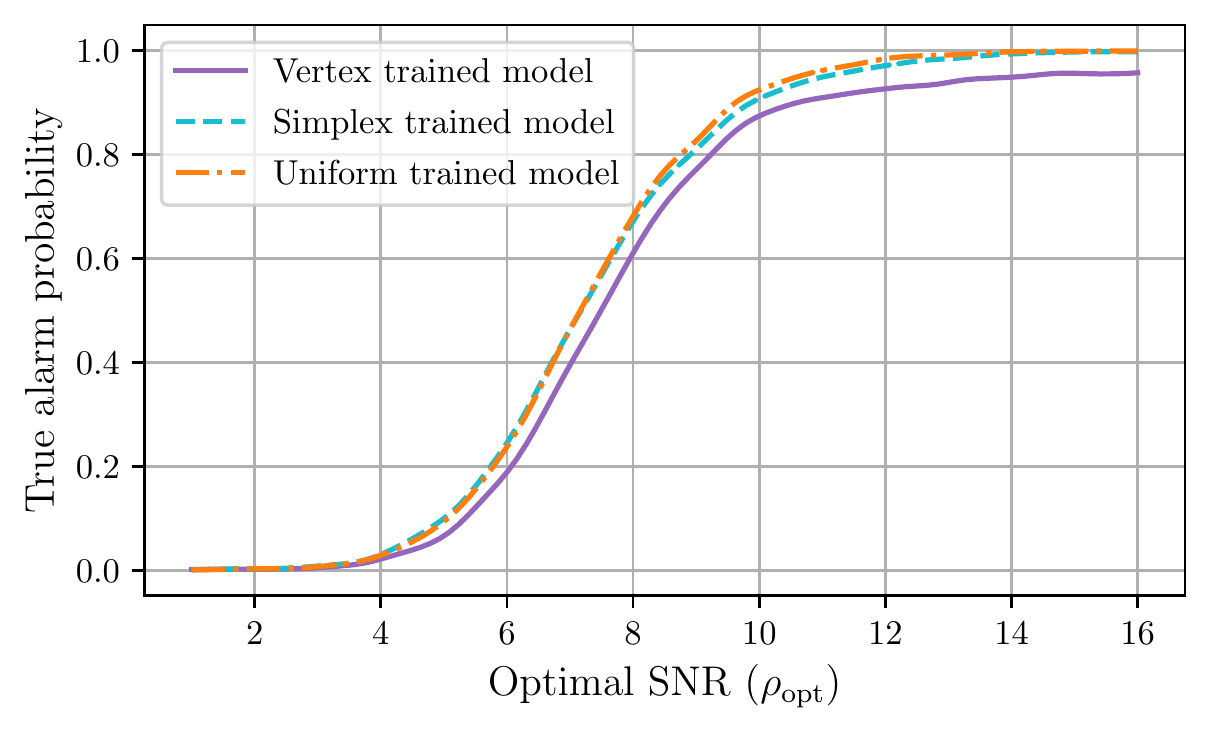}
\caption{Tested on simplex}
\end{subfigure}

\begin{subfigure}{1\textwidth}
\centering
\includegraphics[scale=0.85]{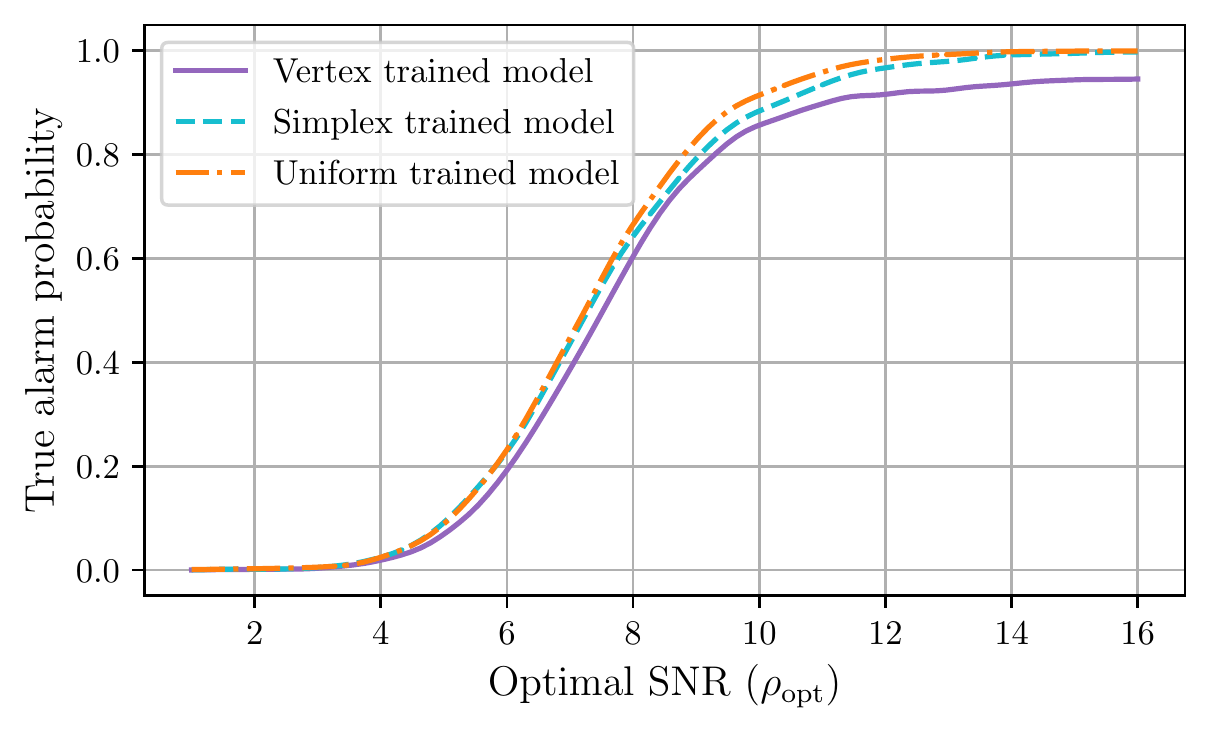}
\caption{Tested on uniform}
\end{subfigure}

\caption{Efficiency curves comparing the performance of the \acp{CNN}. The true alarm probability is plotted as a function of the optimal \ac{SNR} of the signals for a false alarm probability of $10^{-3}$. Each plot shows the performance of a \ac{CNN} trained on vertex, simplex and uniform datasets tested on vertex (a), simplex (b), uniform (c).}
\label{fig:eff_curves}
\end{figure}

%
We now compare the \ac{CNN} results between the datasets by first training three \acp{CNN} on the vertex, simplex, and uniform datasets and then using these models to make predictions on the other testing data that is unseen during the network training process. We compare results for the different permutations in \cref{fig:eff_curves}. In this figure the top panel presents results for the three different networks tested on the vertex data and shows that each model confidently detects all the signals with $\text{\acp{SNR}} >13$. At lower \acp{SNR} the vertex and uniform datasets perform similarly, however, at $\rho_{\text{opt}}\sim 10$ the simplex trained model has slightly worse performance, dropping in true alarm probability by a few percent.

We would expect that when the vertex trained model is tested on vertex data that it outperforms the alternatively trained networks. This is because the vertex data is a subset of each of the other 2 datasets and the network is not required to classify any samples unlike those it has trained on. We also expect that all vertex testing signals should be correctly classified at high \ac{SNR} since the vertex data is a subset of the uniform and simplex training tests. The weaker performance of the simplex trained model could be attributed to the lower density of training signal locations in close proximity to the vertices.   

%
The second panel of \cref{fig:eff_curves} shows the results of the differently trained \acp{CNN} tested on simplex data. As expected the simplex and uniform models detect 100\% of the signals at higher \acp{SNR}. However, the vertex trained model fails to detect all the simplex signals, achieving only $96\%$ true alarm probability at the highest simulated \ac{SNR} $\rho_{\text{opt}}=16$. This is explained when we consider that the simplex data is a subset of uniform data while the vertex data is not. It is interesting to note that the simplex and uniform trained models perform identically (within statistical uncertainty). The uniform model has a larger signal parameter space volume and we might expect it to be more more susceptible to misidentifying instances of the Gaussian noise model as signals from the uniform dataset.    

%
The final panel of~\cref{fig:eff_curves} tests the models on uniform data and again shows that at high \acp{SNR} both simplex and uniform trained models are result in 100\% true alarm probability. One might not expect this since the simplex training data is only a subset of the uniform testing data parameter space. The simplex trained \ac{CNN} in the high \ac{SNR} limit is able to confidently generalise to be able to identify signals from the uniform testing dataset. This is not the case for the vertex trained model which achieves only a 95\% true alarm probability in the high \ac{SNR} limit. The vertex trained \ac{CNN} is not able to fully generalise and identify signal from noise for signals within the class space hyper-cube, nor from within the class space simplex hyper-surface. We also note that specifically in the $\rho_{\text{opt}}\sim 10$ region we see marginally more sensitive results for the uniform trained model when applied to the uniform testing data in comparison to the simplex trained model. This is expected since again the simplex data space is a subset of the uniform data space and the uniform trained model will have explicitly learned how to identify signals in regions distant from the simplex hyper-surface. The simplex trained model performs well despite having to extrapolate away from its training space.        

%
The tests discussed above show that the \ac{CNN} trained on the vertex model only manages full detection when tested on vertex model data. The uniform model performs best in all cases and since it contains signals from the vertex and simplex samples and does not appear to suffer from an increased false alarm probability due to its larger parameter space volume. This suggests that the uniform method of sampling the class space for training or characterising a search algorithm is the most robust and sensitive approach given the intrinsically unknown nature of GW burst signals. Furthermore, since the uniform trained model performs equally as well as the vertex trained model when applied to  vertex test data, we can conclude that the inclusion of the unmodelled signals does not negatively affect the model's performance on modelled signals.

\section{Conclusions}\label{conclusions}

%
%
In this work we present the potential of \acp{GAN} for
burst GW generation. We have shown that \acp{GAN} have the ability to generate plausible time-series burst data and present a novel approach to generating unmodelled waveforms. We have shown that our implementation of a \ac{CGAN} is able to generate five distinct classes of burst like signals through conditional training which can then be utilised for specified signal generations. The \acp{CGAN} allows us to map the parameter space of each signal class into a common abstract latent space in which common signal characteristics are grouped into smoothly connected regions. We are then able to sample from this space as input to the generator network and produce high fidelity random examples of any of our trained signal classes.      

%
Whilst we have trained our \ac{CGAN} on 5 discrete signal classes, each having its own signal parameter space, we have shown that we can subsequently sample from the continuous class space to generate hybrid burst waveforms. This novel aspect of our analysis takes advantage of the learned mapping between individual discrete signal classes. When coupled with the latent space, we are then able to generate hybrid waveforms that span the variation between signal classes and the variation within each class. The resultant hybrid waveforms then represent a generalised set of potential GW burst waveforms that are vastly different from the limited training set. Such waveforms are in demand in GW astronomy as they allow burst search pipeline developers to test and enhance their detection schemes.    

%
To provide a practical example of the usage of these waveforms we have concluded our analysis with a simple search for signals in additive Gaussian noise. We have suggested 3 variations of how to sample from the \ac{CGAN} signal class space and have trained a basic \ac{CNN} separately on those data in order to classify whether a signal was present in the noisy data versus only Gaussian noise. The resulting trained networks were then tested on independent datasets from each of the three signal hybrid classes. The resulting efficiency curves compare the detection sensitivities of the \ac{CNN} as a function of \ac{SNR} and allow us to conclude that in this simple analysis, training the search using the most general set of hybrid waveforms (our ``uniform" set) provides the most sensitive overall result. 

%
In contrast to typical approaches in signal generation this is the first time a \ac{GAN} has been used for generating GW burst data. Our approach allows us to explicitly control the mixing of different signal training classes but the variation within the space of signal properties is determined randomly through sampling of the abstract latent space. In the future, as development in \acp{GAN} and generative machine learning advances it is expected that we will gain greater control over targeted generation of signal features. It will also be important to extend our models to train on, and generate, longer duration waveforms, higher sampling rates, and to be conditioned on additional classes. One such set of additional classes of interest would be the population of detector ``glitches''. These are typically high-amplitude short-duration events in the output of GW detectors that represent sources of terrestrial detector noise rather than that of astrophysical origin. Using a \ac{GAN} to model these would provide us with a tool to simulate an unlimited set of glitches which could be used to better understand their origin and guide us towards more effective methods of mitigation and removal from the data stream. 

%
Another waveform class of interest are those of Supernovae, for which some of our hybrid \ac{GAN} generated waveforms share common features (see Figs.~\ref{fig:simplexd_samples} and~\ref{fig:uniform_samples}). Since Supernovae simulations are extremely computationally costly their are relatively few $\mathcal{O}(100s)$ waveforms available for training. This makes \acp{GAN} an attractive prospect for generating entirely new pseudo-realistic waveform realisations consistent with the prior distribution defined by the training set. The conditional aspect of our \ac{GAN} implementation could also allow the user to specify particular desired physical properties of the generated waveforms. For this Supernovae application specifically, we mention the benefit of extending our current method beyond modelling only a single polarisation.  

%
Having the ability to quickly generate new waveforms is essential to test current GW burst detection schemes~\cite{drago2020coherent,Klimenko_2008, Aso_2008}. They can be used to truly assess their sensitivity to unmodeled sources and identify signal features to which they are susceptible. 

\ack

The authors also gratefully acknowledge the Science and Technology Facilities Council of the United Kingdom. JM is supported by the Science and Technology Facilities Council Newton-Bhabha ST/R001928/1 and the Gravitational-wave Excellence through Alliance Training (GrEAT) Network with China ST/R002770/1 funds. CM and ISH are supported by the Science and Technology Research Council (grant No. ST/ L000946/1) and acknowledge the European Cooperation in Science and Technology (COST) action CA17137. MJW is supported by the Science and Technology Facilities Council [2285031].

\clearpage

\section*{References}
\bibliography{main}

\clearpage

\appendix
\section{List of hyperparameters}

\begin{table}[hb]
\centering
\caption{The architecture and hyperparameters describing our \ac{GAN} consisting of discriminator and generator convolution neural networks. The discriminator casts the class input through a fully connected layer such that its dimensions match the signals input which it then concatenates channel-wise. This is then downsampled through four convolutional layers all activated by Leaky ReLU functions and drops half of the connections at the end of each of these layers. The vector is then flattened to one dimension before fully connecting to a single neuron and its output activated by sigmoid to represent the probability the signal came from the training set. The generator concatenates the latent and class input vectors which is fed to a fully connected layer. This layer is then upsampled by four transposed convolutions. Batch normalisation is applied to the output of the first layer and all convolutional layers are activated by ReLU with the exception of the final layer which is Linear. Finally, the extra dimension introduced for the convolution is removed.}
\resizebox{0.8\textwidth}{!}{%
\begin{tabular*}{\textwidth}{c @{\extracolsep{\fill}} c c c c c c}
\br
\mr
&& Discriminator &&& \\
\mr
Operation & Output shape & Kernel size & Stride & Dropout & Activation \\
Class Input & (5) & - & - & 0  & - \\
Dense & (1024) & - & - & 0 & - \\
Signal Input & (1024) & - & - & 0 &  - \\
Concatenate & (1024, 2) & - & - & 0 &  - \\
Convolutional & (512, 64) & 14 & 2 & 0.5 & Leaky ReLU \\
Convolutional & (256, 128) & 14 & 2 & 0.5 &  Leaky ReLU \\
Convolutional & (128, 256) & 14 & 2 & 0.5 & Leaky ReLU \\
Convolutional & (64, 512) & 14 & 2 & 0.5 &  Leaky ReLU \\
Flatten & (32768) & - & - & 0 &  - \\
Dense & (1) & - & - & 0 & sigmoid \\
\mr
&& Generator &&& \\
\mr
Operation & Output shape & Kernel size & Stride & BN & Activation \\
Class Input & (5) & - & - & \ding{55}  & - \\
Latent Input  & (100) & - & - & \ding{55} & - \\
Concatenate & (105) & - & - & \ding{55} &  - \\
Dense & (32768) & - & - & \ding{55} &  ReLU \\
Reshape & (64, 512) & - & - & \ding{55} & - \\
Transposed Conv & (128, 256) & 18 & 2 & \ding{51} & ReLU \\
Transposed Conv & (256, 128) & 18 & 2 & \ding{55} &  ReLU \\
Transposed Conv & (512, 264) & 18 & 2 & \ding{55} & ReLU \\
Transposed Conv & (1024, 1) & 18 & 2 & \ding{55} & Linear \\
Reshape & (1024) & - & - & \ding{55} & - \\

\end{tabular*}}
\resizebox{0.8\textwidth}{!}{%
\begin{tabular*}{\textwidth}{@{} l l l l l l}
\mr
 Optimizer & Adam($\alpha$ = 0.0002, $\beta_{1}$ = 0.5) \\
 Batch size & 512  \\
 Epochs & 500  \\
 Loss & Binary cross-entropy \\
 \mr
 \br
\end{tabular*}}\\
\label{Tab:gan_training_parms}
\end{table}

\clearpage

\begin{table}[hb]
\centering
\caption{The architecture and hyperparameters describing our CNN consists of four convolutional layers followed by two dense layers. The convolutional and dense layers are activated by the swish function \cite{ramachandran2017searching} and dropout is applied, while the final layer uses the sigmoid activation. The network is trained by minimising the binary cross entropy and optimised with Adam with learning rate $10^{-3}$. We train for 100 epochs with a batch size of 1000.}
\resizebox{0.8\textwidth}{!}{%
\begin{tabular*}{\textwidth}{c @{\extracolsep{\fill}} c c c c c c}
\br
Operation & Output shape & Kernel size & Stride & Dropout & Activation \\
\mr
Input & (1024, 2) & -  & - & - & - \\
Convolutional & (512, 8) & 5 & 2 & 0 & Swish  \\
Convolutional & (256, 8) & 5 & 2 & 0 & Swish  \\
Convolutional & (128, 8) & 5 & 2 & 0 & Swish  \\
Convolutional & (64, 8) & 5 & 2 & 0 & Swish  \\
Dense & (100) & 100 & - & 0.2 & Swish  \\
Dense & (1) & 1 & - & 0 & Sigmoid \\
\end{tabular*}}\\
\resizebox{0.8\textwidth}{!}{%
\begin{tabular*}{\textwidth}{@{}l l l l l l}
\mr
Optimizer & Adam($\alpha$ = 0.001, $\beta_{1}$ = 0.5) & & & & \\
Batch size & 1000 & & & & \\
Epochs & 100 & & & & \\
Loss & Binary cross-entropy & & & & \\
 \br
\end{tabular*}}\\
\label{Tab:cnn_training_parms}
\end{table}

\end{document}